%% file: main.tex
\documentclass[%
 reprint,
superscriptaddress,
 amsmath,amssymb,
 aip,
 super,
]{revtex4-1}

\usepackage{graphicx}
\usepackage{dcolumn}
\usepackage{bm}
\usepackage{enumitem}

\usepackage[utf8]{inputenc}
\usepackage[T1]{fontenc}
\usepackage{etoolbox}

\usepackage{nicefrac}
\usepackage{float}
\usepackage{textcomp}
\usepackage{makeidx}
\usepackage{natbib}
\usepackage{braket}
\usepackage[title]{appendix}
\usepackage{graphics}
\usepackage{physics}
\usepackage[version=4]{mhchem}
\usepackage{wrapfig}
\usepackage{tikz}
\usetikzlibrary{graphs}
\usepackage{pgfplots}
\usepackage{picture}
\usepackage[caption = false]{subfig}
\usepackage{placeins}
\usepackage{booktabs}
\usepackage{siunitx}
\usepackage{amssymb}
\usepackage{amsfonts}
\usepackage{mathrsfs}
\usepackage{mathbbol}
\pgfplotsset{compat=1.16}

\makeatletter
\def\@email#1#2{%
 \endgroup
 \patchcmd{\titleblock@produce}
  {\frontmatter@RRAPformat}
  {\frontmatter@RRAPformat{\produce@RRAP{*#1\href{mailto:#2}{#2}}}\frontmatter@RRAPformat}
  {}{}
}%
\makeatother

\begin{document}

\input{commands}

\title{The microscopic origin of anomalous properties of ice relies on the strong quantum anharmonic regime of atomic vibrations}
\author{Marco Cherubini}
\email{marco.cherubini@uniroma1.it}
\affiliation{Dipartimento di Fisica, Università di Roma Sapienza, Piazzale Aldo Moro 5, I-00185
Roma, Italy}
\affiliation{Center for Life NanoScience, Istituto Italiano di Tecnologia, viale Regina Elena 291, 00161 Rome, Italy}
\author{Lorenzo Monacelli}
\email{lorenzo.monacelli@roma1.infn.it}
\affiliation{Dipartimento di Fisica, Università di Roma Sapienza, Piazzale Aldo Moro 5, I-00185
Roma, Italy}
\author{Francesco Mauri}
\email{francesco.mauri@uniroma1.it}
\affiliation{Dipartimento di Fisica, Università di Roma Sapienza, Piazzale Aldo Moro 5, I-00185
Roma, Italy}

\date{\today}

\begin{abstract}
Water ice is a unique material presenting intriguing physical properties, like negative thermal expansion and anomalous volume isotope effect (VIE). They arise from the interplay between weak hydrogen bonds and nuclear quantum fluctuations, making theoretical calculations challenging. 
Here, we employ the stochastic self-consistent harmonic approximation (SSCHA) to investigate how thermal and quantum fluctuations affect the physical properties of ice XI \emph{ab initio}.
Regarding the anomalous VIE, our work reveals that quantum effects on hydrogen are so strong to be in a nonlinear regime: when progressively increasing the mass of hydrogen from protium to infinity (classical limit), the volume firstly expands and then contracts, with a maximum slightly above the mass of tritium. 
We observe an anharmonic renormalization of about $10\%$ in the bending and stretching phonon frequencies probed in IR and Raman experiments.
For the first time, we report an accurate comparison of the low energy phonon dispersion with the experimental data, possible only thanks to high-level accuracy in the electronic correlation and nuclear quantum and thermal fluctuations, paving the way for the study of thermal transport in ice from first principles and the simulation of ice under pressure.
\end{abstract}

\maketitle
\section{Introduction}

Water is essential for life. It is omnipresent on Earth in all the states of matter, influencing the climate \cite{BartelsRausch2012}, refrigeration and transportation system.  Ice is a molecular crystal composed of individual water molecules held to one another by hydrogen bonds, whose importance ranges from biology to astrophysics. 
Ice manifests polymorphism, typical of molecular crystals. It has been found in 17 different crystalline structure phases to date, embodying the most complex phase-diagram known in literature\cite{Fortes2004,Leadbetter1985,Lobban1998,Kamb1964,Kamb1967,Kamb1968,Kamb1964_2,Engelhardt1981,DOWELL1960}. In each phase, the oxygen atoms are long-range ordered in a specific symmetry with the hydrogen atoms arranged around the oxygen according to the Bernal-Fowler ice rules \cite{Bernal1933}.
Out of the 17 ice phases, some are proton-disordered while the others are proton-ordered.
In this work, we focus on the XI phase \cite{Leadbetter1985}, that is the proton-ordered counterpart of the ordinary ice $I_\text{h}$  \cite{Fortes2004}, stable below T = 72 K, discovered by calorimetric measurements on $\ce{KOH}$ doped ice $I_\text{h}$ \cite{Tajima1982,Tajima1984,Kwada1989}. This makes ice XI a prototype system for phase $I_\text{h}$ . Ice XI has an orthorombic structure with space group $Cmc2_1$ .

Liquid and crystalline water have been extensively studied in recent decades. Besides its pivotal role in biological processes, researchers focused on investigating water thanks to its anomalous properties attracting scientific attention and making theoretical predictions challenging. The great difference in strength between the intermolecular hydrogen bonds and the intra-molecular covalent OH bonds produces a vibrational spectrum with a wide energy range, from low energy rotons to high energy vibrons \cite{Marchi1986,Tse1984,Bosi1973}. 
The stretching vibrational modes of the water molecule have an energy of 3400 $\si{\centi\metre^{-1}}$, which needs a temperature of 4900 K to populate the first excited state. Therefore, the nuclear motion is completely quantum mechanical at room temperature.
Anharmonic effects play a key role in ice, determining, for example, its anomalous thermal expansion \cite{Rttger1994,Fortes2018,Tanaka1998} at low temperature, the inverse volume isotope effect (VIE) \cite{Umemoto2015,Pamuk2012,Salim2016} and the shifts in the vibrational spectra.
The thermodynamic properties of ice have been investigated by using several approximations.
Force-field and first-principles based path integral molecular dynamics (PIMD) and quasi-harmonic approximation (QHA) \cite{Pamuk2012,Ramrez2012,Herrero2011,2016APS..MARK47006F,Umemoto2015,Salim2016} have been employed to describe negative thermal expansion and VIE, enlightening the strong dependence of the results on the force field or the DFT functional used as well as some discrepancies of the QHA at high temperatures and the improvement in the simulations of water and ice obtained by including long-range van der Waals interactions \cite{Wang2011,Pamuk2015,2016APS..MARC20006F,2012APS..MAR.H6001F}.
The vibrational properties of ice have been widely investigated by using mainly Raman, Infrared and neutron spectroscopy \cite{Li2010,Li2010_2,Wjcik2002,Bergren1982,Rice1983,Bertie1964,Scherer1977,Clapp1995,Zasetsky2005,Abe2000,Abe2011,Shigenari2012,Li1995,Arakawa2009,Fukazawa1998}. Alongside experimental works, several theoretical studies focused on the librational modes \cite{Itoh1996,Itoh1998,Erba2009,Gug2015} and on the OH stretching bands \cite{Iftimie2005,Chen2008}, where the authors were able to compute linewidths in good agreement with experimental data but shifted peak positions.
Low energy modes computed with standard \emph{ab initio} techniques display a severe disagreement with experiments. This prevents the study of thermal transport properties in ice \emph{ab initio} and its characterization under pressure, a widely experimentally probed field  \cite{Klotz_2005,Klotz2002,Klotz2005,Strssle2004,Strssle2007,Besson1997,Nelmes2006}.
An accurate description of atomic vibrations is of paramount importance to reproduce thermodynamic and dynamical properties.
In this work, we overcome the intrinsic limitations of other methodologies by using the self-consistent harmonic approximation (SSCHA) \cite{Errea2014,Bianco2017,Monacelli2018,Monacelli2021,Monacelli2021} 
that exploits a full-quantum variational principle on the free energy to include the effect of anharmonicity introduced by thermal and quantum fluctuations.

In Sec. \ref{sec:Theory}  we revise the methodology we employed in the simulations. In Sec.~\ref{subsec:Thermodynamic_results}, we report the temperature evolution of volume (thermal expansion), internal geometry, and bulk modulus. We analyze the volume as a function of isotope mass in Sec.~\ref{sec:VIE} (VIE). Sec.~\ref{subsec:dispersion_results}, \ref{subsec:spectral} and \ref{sec:Dynamical_results} discuss phonon dispersion, overtones and combination modes in the phonon spectrum, and the Infrared, Raman vibrational spectrum of ice, respectively.
In Sec.~\ref{sec:conclusion}, we resume the results and draw the conclusions. 

\section{Methods}
\label{sec:Theory}
We work within the Born-Oppenheimer (BO) approximation \cite{Born1927} to separate electronic and nuclear degrees of freedom. The total electronic energy at fixed nuclei is computed with a Neural Network Potential (NNP) devised in \cite{Cheng2019}, trained on the revPBE0 \cite{Zhang1998,Adamo1999,Goerigk2011} functional with Grimme D3 dispersion correction \cite{Grimme2010,Goerigk2011} to properly account for long-range van der Waals interactions.
\par We solve the nuclear BO Hamiltonian with two different approximations: the quasi-harmonic approximation (QHA) and the self-consistent harmonic approximation (SSCHA).
\par In the QHA, the BO energy surface is expanded as a quadratic function around its minimum at each volume. The free energy is the sum of the BO energy $V(\rscha ,\lbrace \vec{a}_i \rbrace) $ at fixed nuclear position $\rscha$ and cell parameteres $\lbrace \vec{a}_i \rbrace$, and the harmonic vibrational contribution:

\begin{equation}
    \label{eq:QHA_definition}
    \mathcal{F}_{\textit{QHA}} (\rscha,\lbrace \vec{a}_i \rbrace) = V( \rscha, \lbrace \vec{a}_i \rbrace) + F_{\textit{vib}} (\rscha,\lbrace \vec{a}_i \rbrace) 
\end{equation}
where
\begin{equation}
\label{eq:QHA_vib}
\begin{split}
F_{\textit{vib}}( \rscha,\lbrace \vec{a}_i \rbrace) = 
\displaystyle \frac{1}{N_{\bq}} &
\sum_{ \bq \in BZ} \sum_{\mu=1}^{3N}  \bigg[ \displaystyle \frac{\hbar \omega_{\bq\mu}^\mathcal{H}(\rscha,\lbrace \vec{a}_i \rbrace)} {2}+ \\
& + \displaystyle \frac{1}{\beta}  \ln(1- e^{-\beta \hbar \omega_{\bq\mu}^\mathcal{H}(\rscha,\lbrace \vec{a}_i \rbrace)})
\bigg].
\end{split}
\end{equation}

Here, $N_{\bq}$ is the number of $\bq$  points in the Brillouin zone, $\beta  = (k_B T)^{-1}$, and $\omega_{\bq\mu}^\mathcal{H}$ are the volume dependent harmonic frequencies in the $\bq$ point for the $\mu$ mode. However, as in the harmonic model, the frequencies do not show any temperature dependence for a fixed volume.

The vector $\rscha$ describes the positions of the $N$ nuclei in the periodic cell (or supercell), while $\lbrace \vec{a}_i \rbrace$ are the unit cell vectors.

In principle, the QHA free energy is obtained by minimizing the functional $\mathcal{F}_{\text{QHA}}$ in Eq. (\ref{eq:QHA_definition}) at fixed volume and temperature. However, this minimization  is computationally expensive for systems with many degrees of freedom like ice, as it requires the calculation of the harmonic phonon frequencies for each value of the nuclear positions.

In this work (as commonly done), the QHA free energy is computed in the minimum $\rscha_0$ of the BO energy $V(\rscha,\lbrace \vec{a}_i \rbrace)$, obtained by relaxing both the internal coordinates and the cell vectors without vibrations at fixed pressure.

We overcome the intrinsic limitations of the QHA by employing a more sophisticated technique.

The self-consistent harmonic approximation (SSCHA) is a quantum variational principle on the free energy, accounting for quantum and anharmonic effects on nuclei in a nonperturbative way. Within the SSCHA, we optimize the quantum density matrix to minimize the free energy, constraining the density matrix $\tilde \rho$ among the most general Gaussians,  uniquely defined by the average atomic positions (centroids) $\rscha$  and the quantum fluctuations around them (force constant matrix) $\phischatrial$. 

The SSCHA free energy is
\begin{equation}
\begin{split}
    \label{eq:SSCHA_free_energy}
    \mathcal{F}_{\textit{SSCHA}}[\rscha,\phischatrial] = \Avgschatrial{V-\Vcal^{\Hschatrial}}
    + F_{\textit{vib}} (\rscha, \mathbf{\phischatrial})
\end{split},
\end{equation}
where $\Vcal^{\Hschatrial}$ is the potential energy for a trial harmonic Hamiltonian $\Hschatrial$ and the vibrational term has the same functional dependence as in Eq. (\ref{eq:QHA_vib}). The average is computed in an ensemble of configurations generated according to the density matrix $\tilde \rho$.

The time-dependent extension of the SSCHA \cite{Bianco2017,Monacelli2021} gives the possibility to compute the dynamical properties (phonon spectra).

The differences between the two approaches rely in three main points:
\begin{enumerate}[label=\alph*)]
    \item{ The centroid position $\rscha$
    
    As previously discussed, we cannot optimize the centroids within the QHA at a reasonable computational cost, so we employ their equilibrium value without vibrations.
    Conversely, the minimization procedure in the SSCHA allow us to completely optimize the geometry at any temperature, including the average nuclear position $\rscha$.}
    \item{The frequencies:
    
    In the QHA approach, they are the harmonic frequencies.
    In the SSCHA framework, frequencies are the eigenvalues of the dynamical matrix ( the force constant matrix divided by the square root of the masses), obtained through the free energy minimization. Thus, they account for anharmonic quantum and thermal fluctuations. }
    \item{The SSCHA accounts explicitly for deviation of the real ionic energy landscape from the Harmonic approximation $\expval{ V-\Vcal^{\Hschatrial}}_{\rho_{\rscha,\mathbf{\phischatrial}}}$ in Eq.~\eqref{eq:SSCHA_free_energy}. The inclusion of this term makes the SCCHA free energy variational with respect to the exact one. Such property is not shared by the QHA free energy
    }
\end{enumerate}

Both the harmonic and the SSCHA calculation are performed in supercells with periodic boundary conditions. 

In Appendix \ref{app:convergence}, we report a detailed discussion about the convergence properties in the SSCHA and QHA. We found that the results are converged for a 3x3x2 supercell in the SSCHA and a 14x14x14 for the QHA.

For each temperature, we estimated the equilibrium volume $\Omega_{\textit{eq}}(T)$ as the one where the pressure, defined as the derivative of the free energy with respect to a strain tensor $\bvareps$,  $P = -\nicefrac{1}{\Omega}\pdv*{\mathcal{F}}{\bvareps}$, vanishes.

\begin{equation}
    \label{eq:equilibrium_volume}
    \Omega_{\textit{eq}}(T) :\; P(\Omega_{eq}(T),T) = 0
\end{equation}

In the SSCHA framework, we have an analytical equation to compute the pressure for each simulation\cite{Monacelli2018}, while we employed the finite difference approach for the QHA.

When an external pressure is applied to a solid, its volume changes relatively; the bulk modulus is how a crystal withstands modifications of volume under pressure.

\begin{equation}
\label{eq:bulk_modulus}
B(T)=-\Omega_\textit{eq}(T) \pdv{P(\Omega,T)}{\Omega} \biggr |_{\Omega_\textit{eq}(T)}
\end{equation}

\section{Results}
\label{sec:results}
We report a detailed investigation of the phase XI of ice, the proton ordered phase of common ice \cite{Tajima1982}, stable below 72K. Quantum anharmonic effects on the nuclei affect the properties of the hydrogen bonds, producing exotic behaviours, like negative thermal expansion or anomalous VIE.
Soft inter molecular hydrogen bonds coexist with harder intra-molecular covalent $\ce{OH}$ bonds producing phonons with a very wide energy range, heavily impacted by anharmonicity.

It is known in literature that the commonly used QHA produces accurate predictions of thermodynamic properties at low temperature ( $T \leq \SI{100}{\kelvin}$), manifesting instead some inaccuracies at higher temperatures. Therefore, we are using both the QHA and the SSCHA, a more sophisticated technique to take into account quantum fluctuations, to simulate the system, comparing their outcomes.

The first section is dedicated to the anomalous thermal expansion of ice, the bulk modulus, and the temperature dependence of the crystalline properties.

\subsection{Thermodynamic properties}
\label{subsec:Thermodynamic_results}

\begin{figure}
    \centering
    \includegraphics[width=\columnwidth]{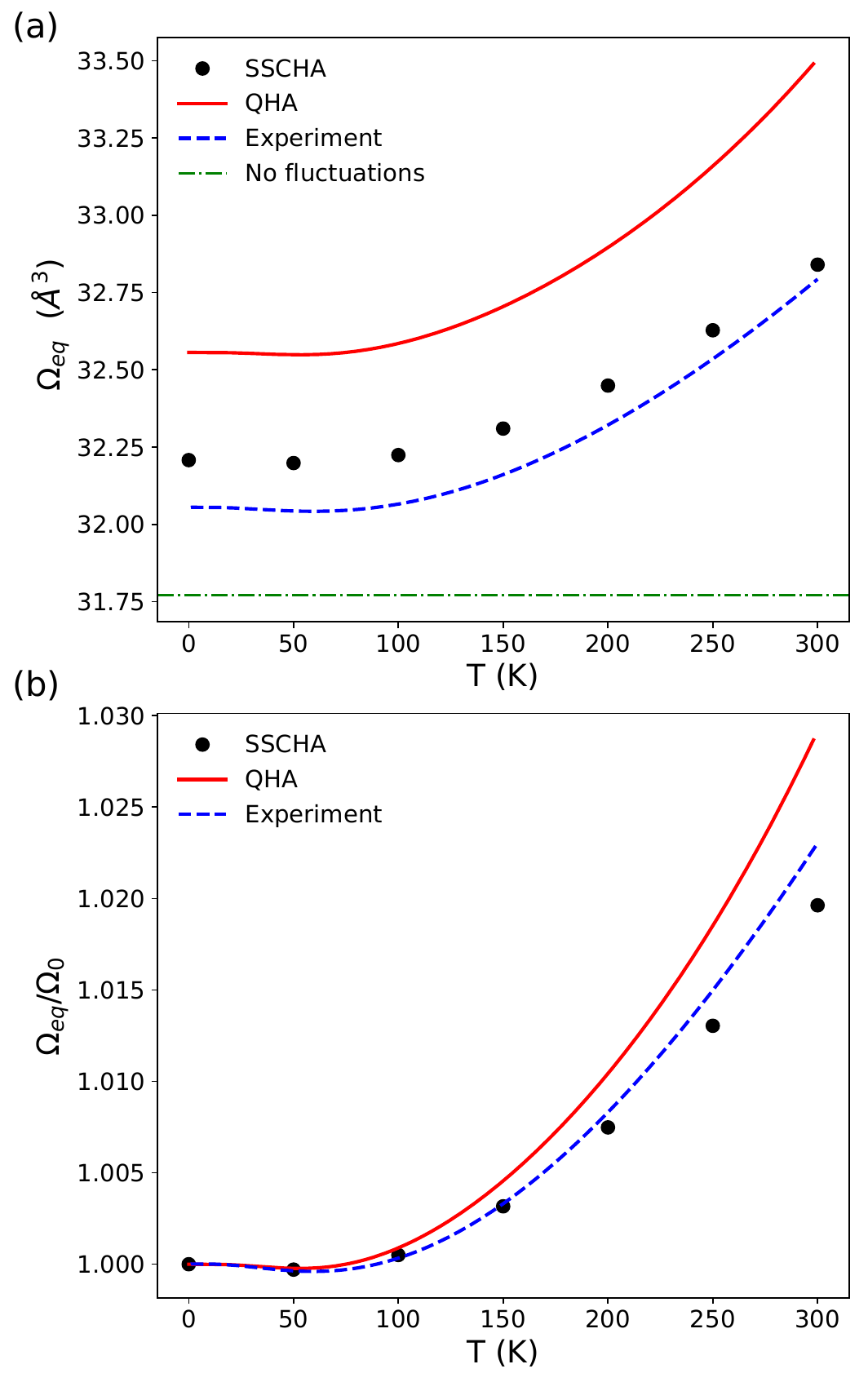}
    \caption{Equilibrium volume per $\ce{H_2O}$ molecule of ice as a function of temperature. $\mathbf{a}$ Comparison of the absolute value of the volume in the SSCHA (black circles) and QHA (solid red line) with the experimental measurements in \cite{Fortes2018} (blue dotted line). The classical equilibrium volume, where thermal and quantum fluctuations are neglected, is the green dot-dashed line. $\mathbf{b}$ The normalized equilibrium volume $\nicefrac{\Omega_\textit{eq}(T)}{\Omega_\textit{eq}(T=0)}$ in the SSCHA (black circles) and QHA (red solid line) is compared to the experiment (blue dotted line).}
    \label{fig:normalized_volume}
\end{figure}

The absolute value of the equilibrium volume per $\ce{H_2O}$ molecule is in Fig. \ref{fig:normalized_volume} ($\mathbf{a}$). We notice a considerable effect of the zero point motion that shifts the curves one respect to the other. The zero-temperature equilibrium volumes predicted by the different theories and their percentage shift with respect to the classical limit are reported in the first part of Table \ref{tab:volumes}. The agreement with the experiment is much better in the SSCHA than in the QHA picture.
This data are discussed in detail in Sec. \ref{subsec:VIE_results}, concerning the isotope volume effect. Since $\Omega_{\textit{eq}}(T=0)$ is theory dependent, to compare the thermal expansion, in Fig. \ref{fig:normalized_volume} ($\mathbf{b}$) we report the normalized volume $\nicefrac{\Omega_{\text{eq}}(T)}{\Omega_{\text{eq}}(T=0)}$.

In the low-temperature regime, $T \leq 50 K$, the predictions for the normalized equilibrium volume in the QHA and SSCHA are very similar and in a good match with the experiment \cite{Fortes2018}. For higher temperatures, essential differences between the two theories arise. The SSCHA agrees with experimental measurements within a 0.1 $\%$ up to \SI{200}{\kelvin}. The QHA, otherwise, deviates from the experimental data above \SI{100}{\kelvin}. This establishes the success of the SSCHA theory, overcoming other state-of-the-art techniques for studying the thermal expansion of ice.

\begin{figure}
    \centering
    \includegraphics[width=\columnwidth]{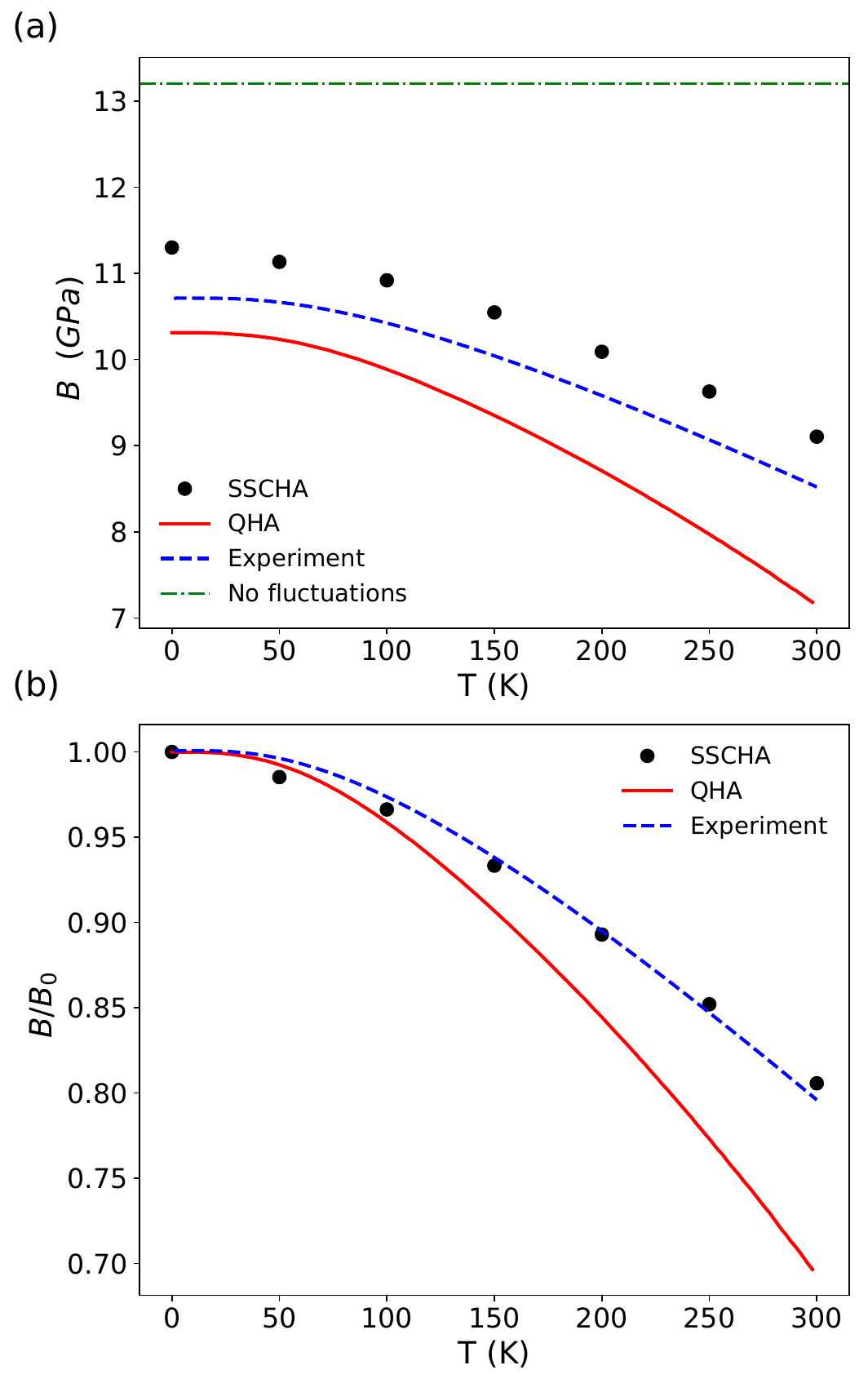}
    \caption{$\mathbf{a}$ Temperature dependence of the bulk modulus. The QHA (red solid line) and SSCHA (black circles) results are compared with the experiment \cite{Fortes2018} (blue dashed line). The green dot-dashed line is the classical value for the bulk modulus, obtained by neglecting thermal and quantum fluctuations. $\mathbf{b}$ Comparison of the normalized bulk modulus $\nicefrac{B(T)}{B(T=0)}$ in QHA and SSCHA with the experiment. The bulk modulus is computed as in Eq. (\ref{eq:bulk_modulus}).}
    \label{fig:bulk}
\end{figure}

Eq. (\ref{eq:bulk_modulus}) suggests a direct dependence between equilibrium volume and bulk modulus. The absolute value and the normalized one $\nicefrac{B(T)}{B(T=0)}$ of the bulk modulus are reported in Fig. \ref{fig:bulk}. Already at $T = \SI{0}{\kelvin}$, the bulk modulus is strongly renormalized by ionic quantum fluctuations by a $ 22 \%$ and $ 15\%$ in the QHA and SSCHA, respectively (Fig.~\ref{fig:bulk} ($\mathbf{a}$) ). 

The bulk modulus has an anomalous strong temperature dependence; experimental data \cite{Fortes2018} show a $20\%$ reduction from 0 to 300K. The SSCHA reproduces this behavior perfectly (Fig.~\ref{fig:bulk} ($\mathbf{b}$) ), while the QHA overestimates the bulk modulus reduction of $10\%$. This strong temperature dependence originates by $64\%$ from volume expansion and the remaining  $ 36\%$ from vibrational free energy. We refer to App. \ref{app:bulk_modulus} for further details. 

This result is fundamental to geophysics, where the compressibility of a solid is of paramount importance for studying the inner composition of Earth \cite{Matas2007,Irving2018}. The combined employment of high accuracy in the electronic exchange-correlation and the accurate description of quantum nuclear motion provided by the SSCHA correctly describe the thermodynamics properties of ice beyond any other simulation performed so far.

\begin{figure}
    \centering
    \includegraphics[width=\columnwidth]{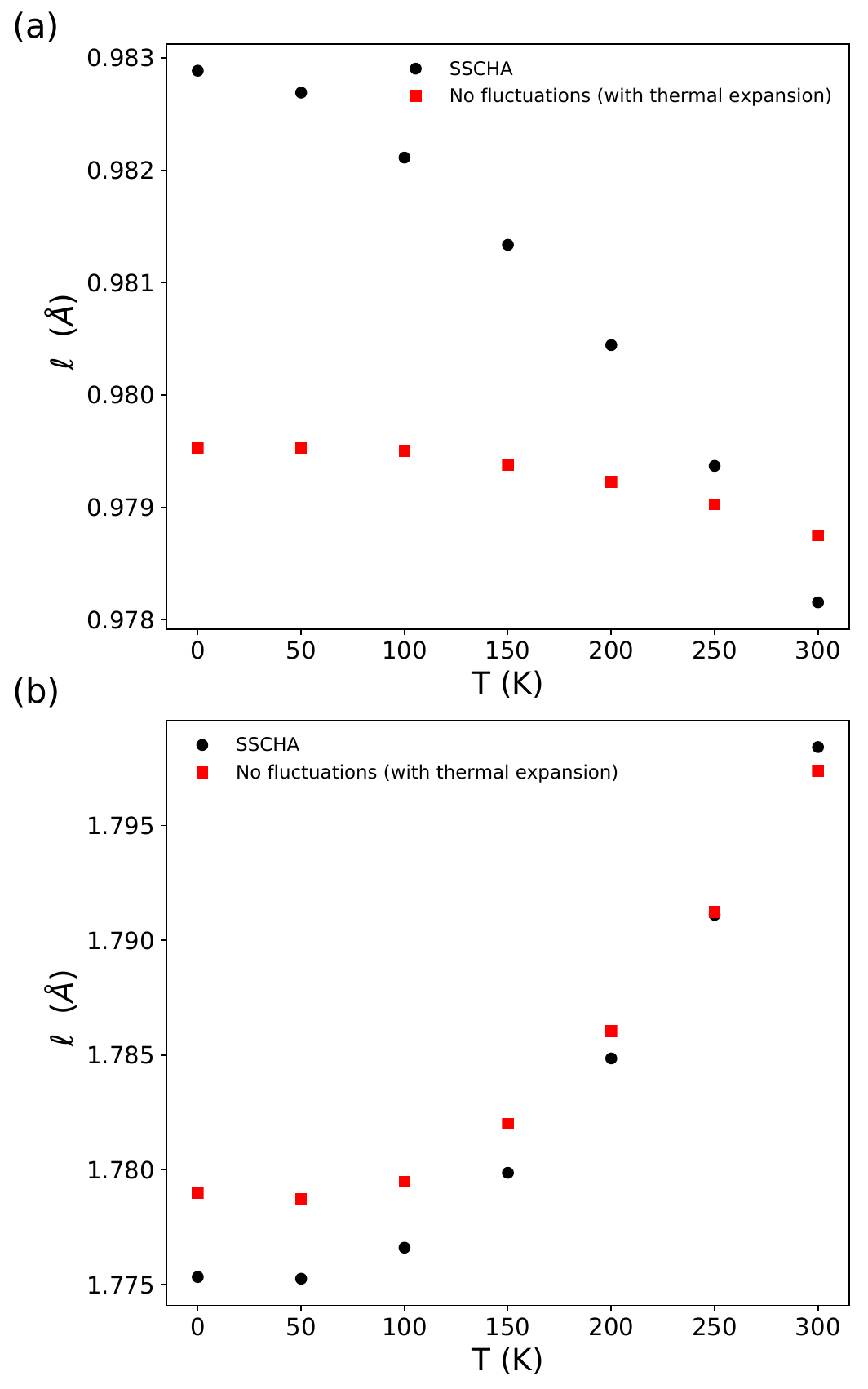}
    \caption{$\mathbf{a:}$ Temperature dependence of the covalent bond length. The SSCHA bond length (black circles) is compared with the classical result (red squares), computed from the minimization of the BO energy $V(\rscha,\lbrace \vec{a}_i \rbrace)$, where, the effect of the thermal expansion is introduced by fixing the volume to be the SSCHA equilibrium one at each temperature, in order to point out the contribution of thermal and quantum fluctuations. $\mathbf{b:}$ Temperature dependence of the hydrogen bond length. Both the SSCHA and the classical bond lengths are computed for the same conditions as in panel $\mathbf{a}$.}
    \label{fig:bond_lengths}
\end{figure}

The SSCHA also provides information about the geometry of ice.
We report the temperature dependence of the average covalent bond length $\ce{OH}$ in Fig. \ref{fig:bond_lengths} ($\mathbf{a})$ and the average hydrogen bond length in Fig. \ref{fig:bond_lengths} ($\mathbf{b}$). 

Counterintuitively, the water molecules shrink upon heating at high temperatures (Fig. \ref{fig:bond_lengths} ($\mathbf{a}$)). 
This is only marginally a consequence of intermolecular hydrogen bond weakening due to the increasing distance between molecules with temperature, but rather a complex effect of anharmonicity triggered by molecular vibration. If we relax the structure with static nuclei at the SSCHA equilibrium volume for each temperature (Fig. \ref{fig:bond_lengths} ($\mathbf{a}$), red squares), we explain only the 15$\%$ of this effect.

Since the vibration that deforms the water molecule is a stretching mode, with $\omega \simeq 3400 \si{\centi\metre^{-1}}$ and excitation temperature $T \simeq 4900K$, this relevant temperature dependence can only be explained by the anharmonic interaction between translational molecular modes, the only ones whose population changes in this temperature range, and the stretching mode, that affects the \ce{OH} bond length.

Conversely, the volume expansion explains thoroughly the widening of the hydrogen bond with temperature (Fig. \ref{fig:bond_lengths} ($\mathbf{a}$)).

\subsection{Volume Isotope Effect}
\label{sec:VIE}

According to classical mechanics, the equilibrium volume does not depend on the mass of the atoms, and, thus, it is isotope independent.
Quantum effects overturn this simple situation. In most crystalline systems, the heavier the isotopes the smaller the equilibrium volume. In rare exceptions, like ice, by substituting hydrogen with deuterium the equilibrium volume increases \cite{Pamuk2012,Umemoto2015}. This is known as anomalous volume isotope effect.

In Table \ref{tab:volumes}, we report the equilibrium volume per water molecule at zero temperature with protium mass of hydrogen, computed without thermal fluctuations, with quantum effects and harmonic phonons (QHA) and with full anharmonic quantum zero point motion (SSCHA) compared with the experiment in Ref. \cite{Fortes2018}.

As seen, the full quantum anharmonic theory is the closest match with the experiment, with an error smaller than 0.5$\%$. The discrepancy between the SSCHA theory and the volume considering the quantum zero point motion but not anharmonicity (QHA) is the same as neglecting ionic motion.

Anharmonicity affects the results in two ways: it changes the average position of nuclei (See Fig. \ref{fig:bond_lengths}) and modifies the vibrational frequencies.  To shed light on which effect dominates the volume expansion, we repeated the harmonic calculation by fixing nuclear positions to the SSCHA result  (QHA @ $\lbrace \rscha \rbrace_{\text{SSCHA}}$) and by employing also the frequencies shifted by the anharmonicity (QHA @ $\lbrace \rscha,\omega \rbrace_{\text{SSCHA}}$).  The results are reported in the second part of Table \ref{tab:volumes}. The last row indicates the volume difference with respect to the full anharmonic theory.

This analysis reveals the origin of the discrepancies between a quasi-harmonic approach and a full non perturbative anharmonic treatment of nuclear vibrations, unveiling how the key role played by anharmonicity is on the frequency renormalization rather than the significative structural changes.

\begin{table}
    \setlength{\tabcolsep}{0pt}
    \centering
    \resizebox{0.5\textwidth}{!}{%
    \begin{tabular*}{0.55\textwidth}{@{\extracolsep{\fill}}*{5}{c c c c c}}
    \toprule
    & Classic  & QHA & SSCHA & Exp. \cite{Fortes2018} \\ 
    \midrule
    $\Omega_{\text{eq}} (\AA^3) $ & 31.771 & 32.555 &32.207&32.055\\
    $\%$   & 0&2.47&1.38&\\
    \bottomrule
    \end{tabular*}}
    \resizebox{0.5\textwidth}{!}{%
    \begin{tabular*}{0.55\textwidth}{@{\extracolsep{\fill}}*{3}{c c c c c}}
    \toprule
    & QHA @ $\lbrace \rscha \rbrace_{\text{SSCHA}}$ &  QHA @ $\lbrace \rscha,\omega \rbrace_{\text{SSCHA}}$\\ 
    \midrule
    $\Omega_{\text{eq}} (\AA ^3) $ & 32.613 & 32.068\\
    $\Delta (\AA^3) $   & 0.406 & -0.139\\
    \bottomrule
    \end{tabular*}}
    \caption{The first part of the table reports the equilibrium volume per water molecule of ice at T=0K. In the first row there is the comparison of the volume computed in the QHA and SSCHA with the experiment \cite{Fortes2018} and with the classical result, where quantum and thermal fluctuations are neglected. In the second row, we report the percentage shift with respect to the classical volume. 
    The second part of the table shows the equilibrium volume per water molecule in the QHA when free energy of Eq. (\ref{eq:QHA_definition}) is computed using the SSCHA equilibrium positions $\lbrace \rscha \rbrace_{\text{SSCHA}}  $ or using both the SSCHA positions and frequencies and their error with respect to the SSCHA result.}
    \label{tab:volumes}
\end{table}

Elucidated the crucial role of anharmonicity in describing the correct volume expansion, we systematically explored the volume effect by varying the mass of the isotopes, both hydrogen and oxygen.

At first, we investigate the dependence of the equilibrium volume at T=0K on the hydrogen mass, numerical simulations give the possibility to modify this parameter without limitations, while experiments are available only for hydrogen and deuterium \cite{Fortes2018,Rttger1994}. 

\begin{figure}
    \centering
    \includegraphics[width=8.6cm]{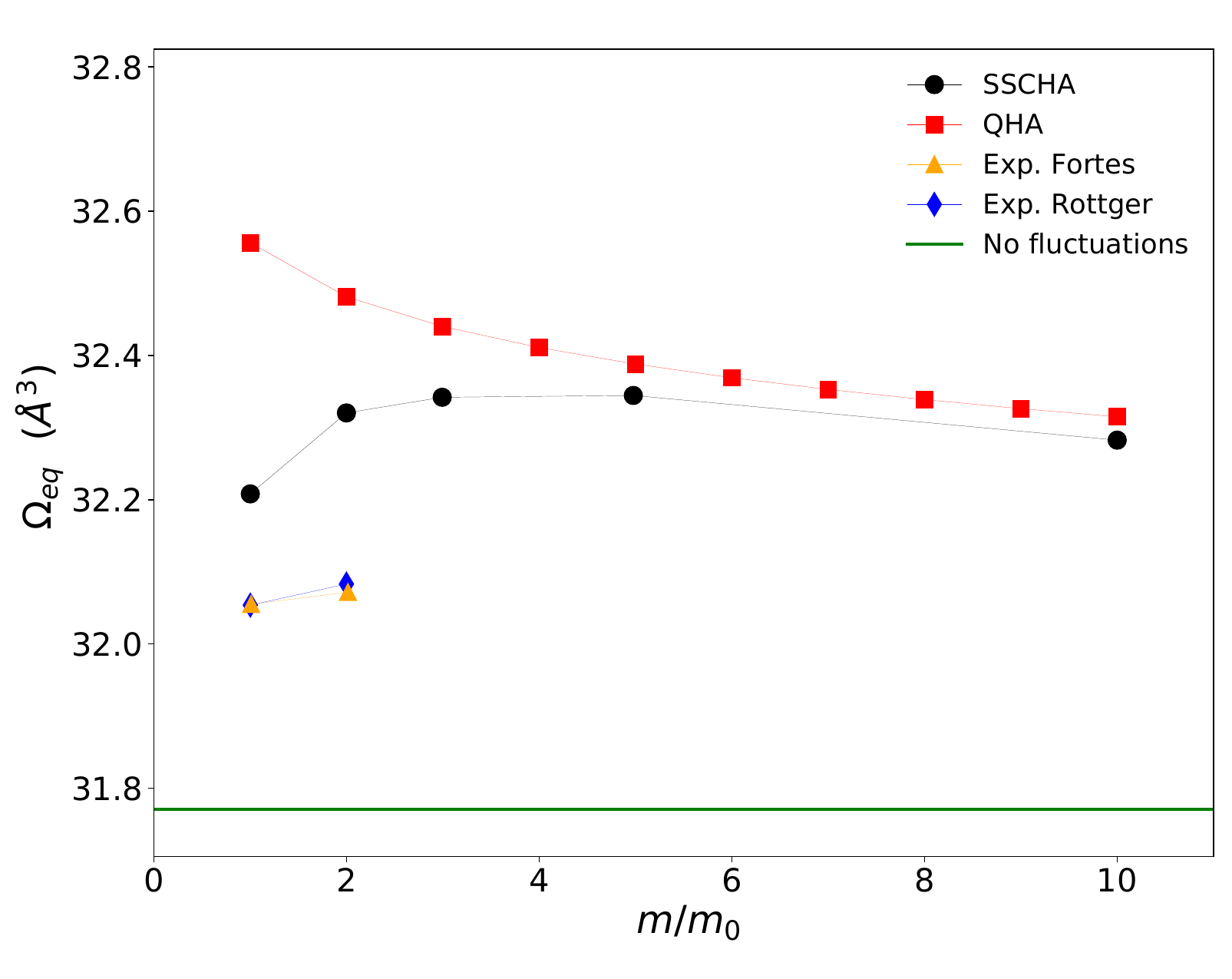}
    \caption{Dependence of the equilibrium volume per water molecule of ice on the hydrogen mass at T=0K. The results obtained in the QHA (red squares) and in the SSCHA (black circles) are compared with two experimental results (yellow triangles and blue diamonds) \cite{Fortes2018,Rttger1994}. The mass independent classical equilibrium volume is reported as a reference (green solid line). } 
    \label{fig:Isotopic}
\end{figure}

In Fig. \ref{fig:Isotopic}, we compare the equilibrium volumes per water molecule obtained at different levels of the theory with two experimental measurements \cite{Rttger1994,Fortes2018}. The continuous line is the classical limit used as a reference.
The numerical values for the volume difference at T = 0K in the theoretical models and the experiments are reported in Table \ref{tab:diff}. 
The complete anharmonic theory (SSCHA) correctly predicts the sign of the VIE, while the quasi-harmonic approach fails, predicting a volume reduction. However, the SSCHA heavily overestimates the experiment, resulting in a volume difference from 3.8 to 6.7 more prominent than the measured data. 

Experimental data are measured on hydrogen-disordered samples of ice $I_h$, while the simulation is performed on the hydrogen-ordered ice XI. To unveil the role of hydrogen ordering, we repeated the calculation of the VIE in a hydrogen-disordered structure for ice $I_h$ with 24 atoms per unit cell. We obtained a result deviating by a 3.5 \% from the hydrogen-ordered structure, unveiling how hydrogen ordering doesn't significantly affect the VIE, and it is not in origin beyond the discrepancy between theory and experiments. 

Our calculation's most relevant source of error is in the electronic correlation: we repeated the simulation employing a different electronic energy engine (a neural network trained on RPBE with Grimme D3 dispersion correction \cite{Cheng2019}). We obtained a result deviating by a $38 \%$ from our simulation, giving a rough estimation of the error introduced by the DFT functional.

The QHA and SSCHA volumes are different for the physical isotopes of hydrogen (protium, deuterium, and tritium). The difference disappears as we increase the hydrogen mass. This derives from the reduced role of anharmonicity for higher mass where quantum fluctuations at zero temperature are smaller. The crossover above which the QHA correctly reproduces the VIE occurs for an isotope mass of five times the hydrogen's one, not a stable isotope.
This means that the quantum regime of protium, deuterium, and tritium is anharmonic, beyond the range of validity of the quasi-harmonic theories.

\begin{figure}
    \centering
    \includegraphics[width=8.6cm]{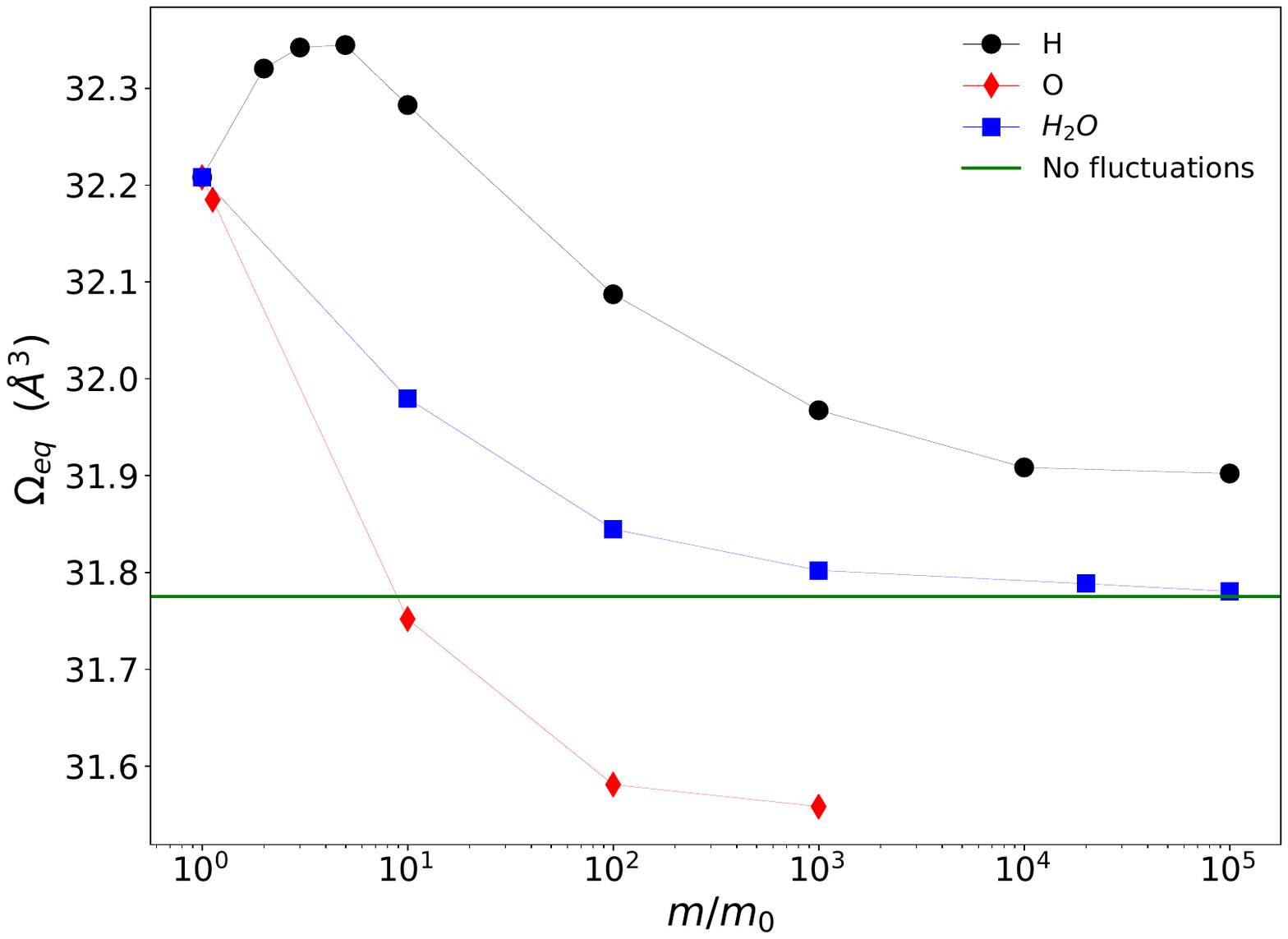}
    \caption{Volume isotope effect for all the atomic species in ice XI in the SSCHA framework at T=0K. Black circles indicate the equilibrium volume per $\ce{H_2O}$ molecule when the hydrogen mass is increased by keeping the oxygen mass fixed to its $\ce{^{16}O}$ isotope value. Red diamonds show the equilibrium volume when the oxygen mass is varied with fixed hydrogen mass. Blue squares stand for the equilibrium volumes when the mass of the entire water molecule is increased until reaching the classical limit shown as a reference (green solid line). }
    \label{fig:Isotopes_SSCHA}
\end{figure}

In Fig. \ref{fig:Isotopes_SSCHA}, we show the equilibrium volume of the solid varying separately the mass of each atomic species (hydrogen and oxygen) until reaching the classical limit of infinite mass, as well as the evolution of the equilibrium volume when the mass of the whole molecule is increased.  Obviously, in nature, only a few of these combinations exist and are stable, but we can infer the quantum nature of each element from this plot.

By increasing the mass of the whole solid, we manage to approach the classical limit (the continuous line) for $\nicefrac{m}{m_{\ce{H2O}}} \gtrsim 10000$.

The largest natural atomic species weighs only 238 times the mass of protium. This unveils how quantum effects on nuclei, usually neglected in atomistic calculations, are of paramount importance even with "heavy" atoms. 
This is further proved by the isotope volume effect of oxygen that, scaled to its much lower mass ratio between its natural isotopes, is bigger than hydrogen.

Fig. \ref{fig:Isotopes_SSCHA} reveals a nonmonotonous volume expansion compared with quantum fluctuations for the hydrogen isotopes. First, the equilibrium volume expands when we increase the mass, reaching the maximum value for $m \simeq 5m_H$, then we observe a contraction to the classical value for bigger masses. This behavior explains that the VIE is due to a crucial nonlinear regime of quantum fluctuations in ice, overturning the hypothesis of a volume reduction due to quantum effects. This exotic behavior cannot be explained in a quasi-harmonic picture, as evident from Fig. \ref{fig:Isotopic}.

\label{subsec:VIE_results}

\begin{table}
    \setlength{\tabcolsep}{0pt}
    \centering
    \resizebox{0.5\textwidth}{!}{%
    \begin{tabular*}{0.55\textwidth}{@{\extracolsep{\fill}}*{6}{c c c c c c c}}
    \toprule
    & SSCHA & QHA &  Exp. \cite{Fortes2018} & Exp. \cite{Rttger1994} & SSCHA RPBE-D3 \\ 
    \midrule
    $\Delta \Omega (\AA^3) $ &0.112&-0.075&0.017&0.029&0.156 \\
    \bottomrule
    \end{tabular*}}
    \caption{Equilibrium volume per water molecule difference $\Omega_{\ce{D_2O}}-\Omega_{\ce{H_2O}}$ at T=0K. The SSCHA and QHA results computed in the converged meshes are compared with two experimental measurements \cite{Fortes2018,Rttger1994}. The last column shows the volume difference computed in the SSCHA by using the NNP RPBE-D3 functional devised in Ref. \cite{Cheng2019} in order to analyze the dependence of the VIE on the functional used. }
    \label{tab:diff}
\end{table}

\subsection{Phonon dispersion}
\label{subsec:dispersion_results}

The coexistence of strong intra-molecular and weak intermolecular bonds in ice produces a vast vibrational spectrum. To compare with experimental results, we computed the real phonons from the dynamical interacting Green function within the time-dependent SSCHA \cite{Bianco2017,Monacelli2021} (TD-SSCHA) to fully account for dynamical quantum anharmonic effects (see App. \ref{app:methods} and \ref{app:dispersion} for further details). We employed the static approximation of the self-energy for the low energy modes, as described in App. \ref{app:methods} and Refs. \cite{Bianco2017,Bianco2018}.

\begin{figure*}
    \centering
    \includegraphics[width=\textwidth]{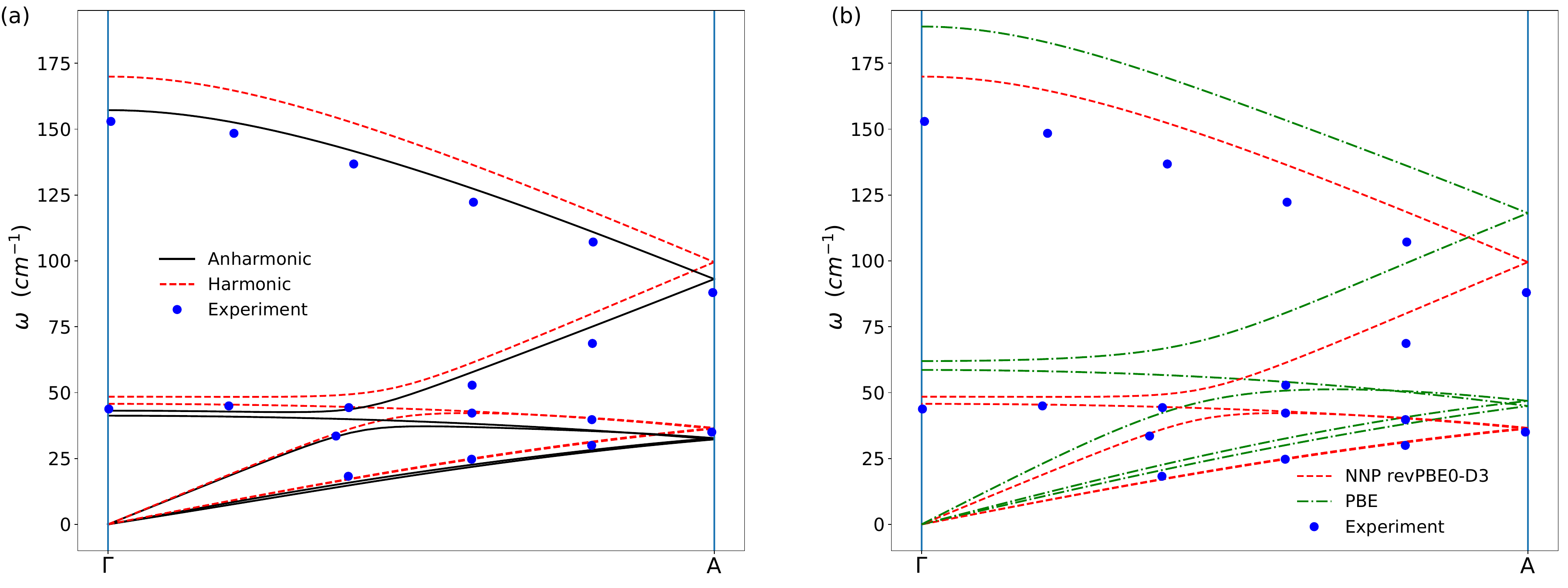}
    \caption{Comparison of the low-energy phonon dispersion (molecular translations) with the experiment \cite{Strssle2004} (blue circles) for deuterated ice at T=140K and P=0.05 GPa. $\mathbf{a}$ Harmonic phonons (red dashed line) and anharmonic phonons (black solid line), computed as the poles of the interacting one-phonon Green function in the static limit are shown. $\mathbf{b}$ Effect of the DFT functional on the harmonic phonon dispersion: The red dashed line is the harmonic dispersion (the same as in panel $\mathbf{a}$) computed with the hybrid revPBE0-D3 functional fitted with the NNP \cite{Cheng2019} (employed in our work for the anharmonic calculations). The green dot dashed lines show the harmonic dispersion calculated with PBE \cite{Perdew1996} functional. }
    \label{fig:Dispersion_exp}
\end{figure*}

We compare the harmonic phonons and the anharmonic (SSCHA) phonons for $\ce{D_2O}$ ice at T=140K and P=0.05 \si{GPa} with the experiment \cite{Strssle2004} in Fig. \ref{fig:Dispersion_exp} ($\mathbf{a}$). Details on the computation are in App. \ref{app:dispersion}.

The harmonic acoustic phonons are in good agreement with the experiment and deviate significantly from experimental data at high frequency ($\omega \ge \SI{120}{\per\centi\meter}$). Anharmonic effects correct the deviation, obtaining a perfect match between theory and experimental data. The harmonic energy of the lowest acoustic branch overestimates the speed of sound of $20\%$, introducing a substantial error in the determination of thermal transport properties, further stressing the fundamental role of anharmonicity in thermal conductivity.

We report an excellent agreement between experimental data and an \emph{ab initio} simulation of ice for the first time. This result has a profound impact enabling the first principles simulation of thermal transport, where an accurate description of the acoustic phonons is required. Moreover, low-energy phonons are the only modes detectable at high pressure; our work paves the way to characterize ice under pressure further.

The astonishingly good agreement we achieve is merit of the combined effect of the correct treatment of anharmonicity and of the electronic functional adopted for the calculation: we report in Fig. \ref{fig:Dispersion_exp} ($\mathbf{b}$) the comparison of harmonic phonons between NN-revPBE0 (employed in our work) and PBE, the common choice in \emph{ab initio} atomistic simulations of water \cite{Pamuk2012,Yoo2009,Rusnak2012}. We notice a considerable dependence on the DFT functional. The use of the NNP improves the PBE harmonic dispersion, where the error committed approximating the experimental points ranges from 30\% to 36\%  e.g. in the A point.

\subsection{Spectral function}
\label{subsec:spectral}

The phonon spectral function $\sigma(q, \omega)$ gives access to the quasiparticles energies and lifetime. The spectral function is proportional to the signal probed in scattering experiments, as neutron or X-ray scattering, and it is computed from the diagonal elements of the dynamical one-phonon Green function $G(q, \omega)$. 

\begin{equation}
    \label{eq:spectral_function}
    \sigma(q, \omega) = -\frac{\omega}{\pi} \Tr \Im G(q,\omega)
\end{equation}

The details for the calculation are reported in App.~\ref{app:compuational_details}.

The phonon density of states (DOS) computed with the SSCHA dynamical matrix at equilibrium (without the self-energy correction) describes anharmonic non-interacting phonons, while the dynamical spectral function encapsulates all the effects of phonon-phonon interactions, where, the addition of the self-energy term (Eq. \ref{eq:self_bubble}) may produce combination of modes (See Eq. \ref{eq:Lambda_tensor} and Eq. \ref{eq:f_function} in App. \ref{app:methods}). 

\begin{figure}
    \centering
    \includegraphics[width=\columnwidth]{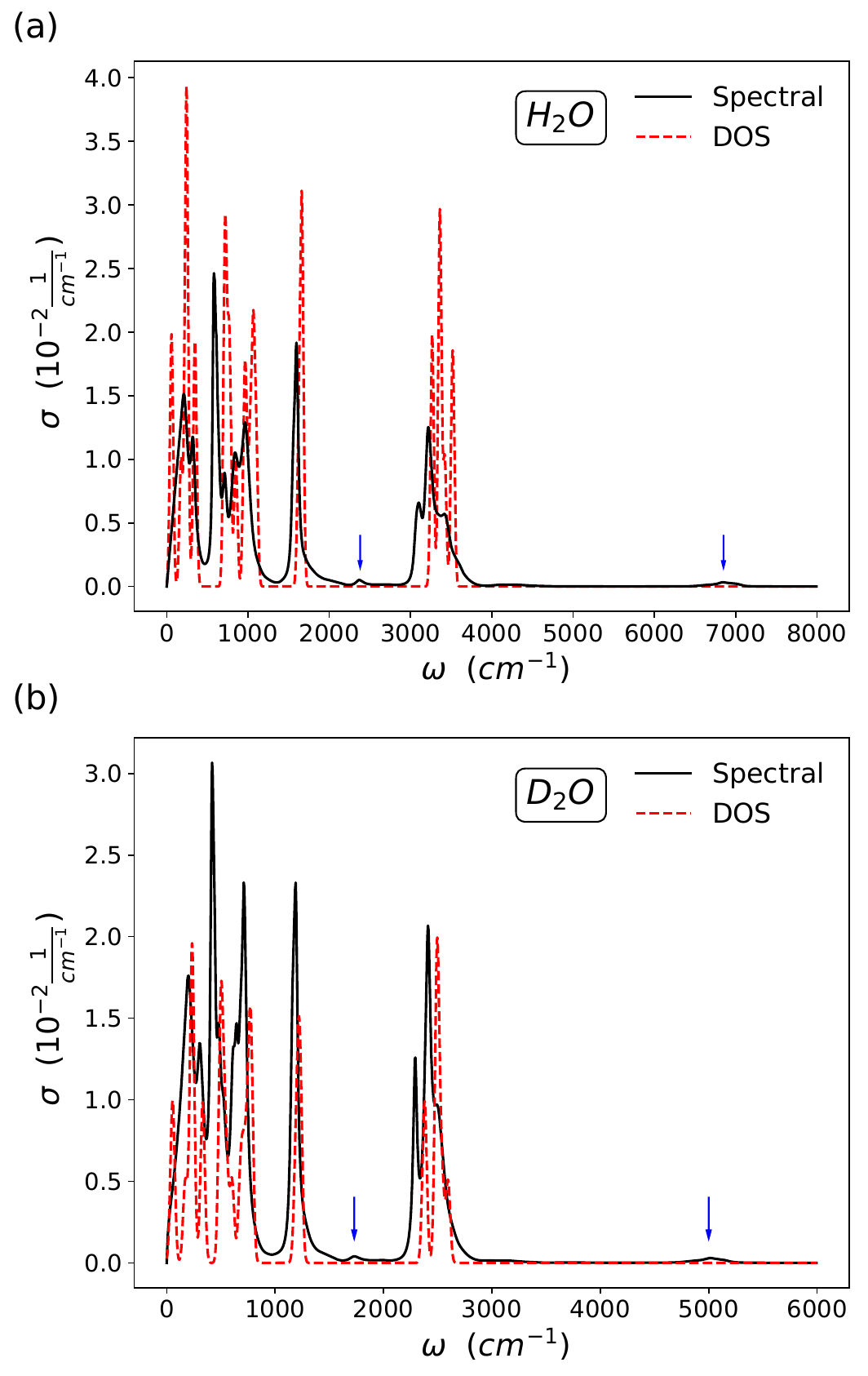}
    \caption{Comparison of the density of states and spectral function at $\Gamma$. $\mathbf{a}$ $\ce{H2O}$ ice at T=200 $\si{K}$ and ambient pressure. Spectral function (black solid line) and DOS (red dashed line) are shown $\mathbf{b}$ $\ce{D2O}$ ice DOS (red dashed line) and spectral function (black solid line) at T=140 $\si{K}$ and P=0.05 $\si{GPa}$. In both panel $\mathbf{a,b}$, the DOS are computed by using the SSCHA dynamical matrix at equilibrium without the inclusion of the self-energy term and by adding an artificial smearing factor of \SI{20}{\per\centi\metre}. Blue arrows indicate the combination modes and anharmonic overtones.}
    \label{fig:dos_combination_modes}
\end{figure}

In Fig. \ref{fig:dos_combination_modes} ($\mathbf{a}$) (\ref{fig:dos_combination_modes} ($\mathbf{b}$)), we report the comparison between the phonon DOS and the spectral function at $\Gamma$ of \ce{H_2O} (\ce{D_2O}) to enlighten the presence of combination modes and anharmonic overtones.
We reveal two structures in the spectral function absent in the DOS: one occurs at energies between the bending and stretching bands, the other at twice the frequency of the stretching modes.

We can dissect the interaction between phonon bands to unveil which modes originate these satellite peaks in the spectral function (see App. \ref{app:methods} for further details).

\begin{figure}
    \centering
    \includegraphics[width=8.6cm]{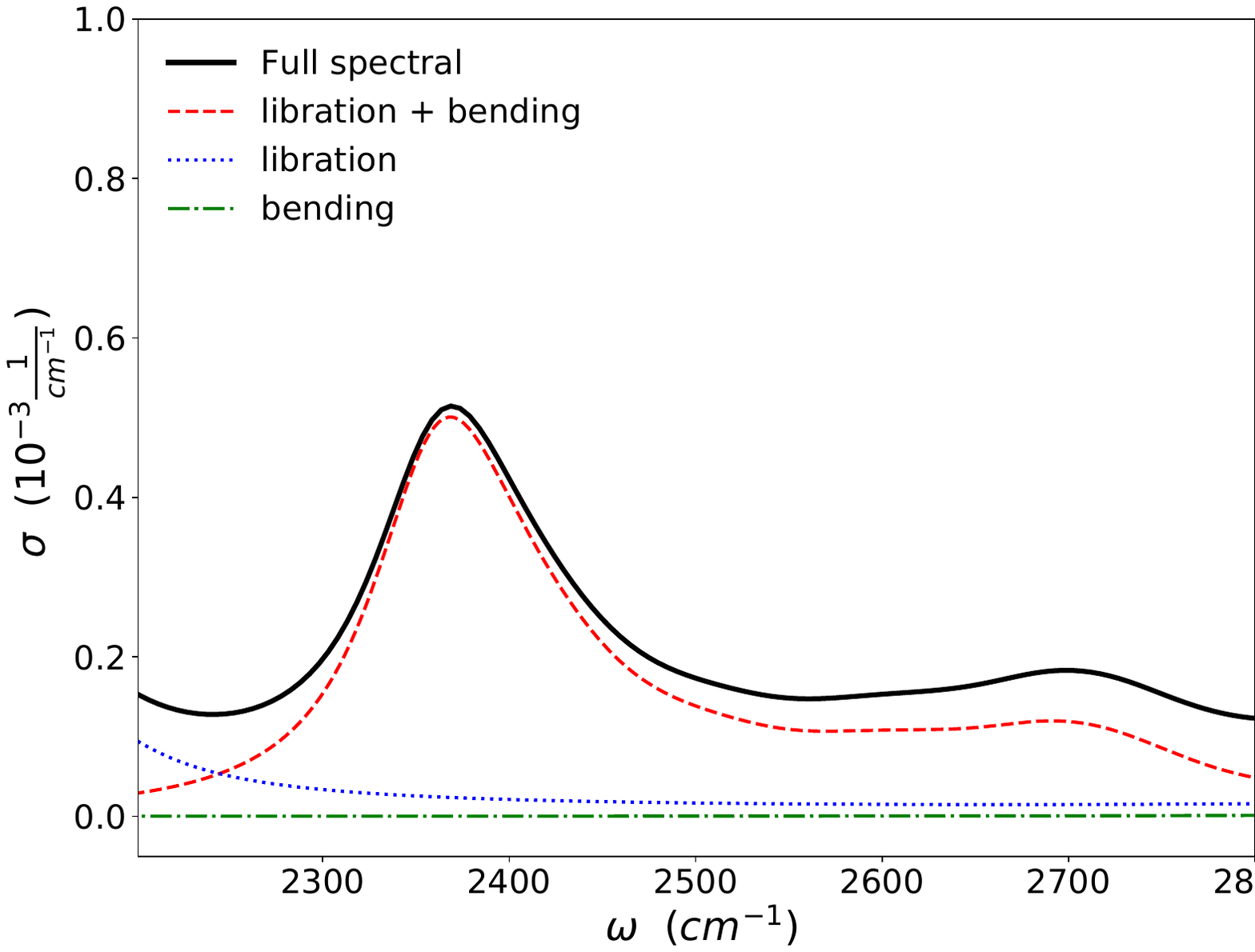}
    \caption{Spectral function of $\ce{H_2O}$ ice at T = 200 $\si{K}$ in the energy range of the first combination mode at $\Gamma$. A comparison between the full spectral function (solid black line) and that obtained by selecting only the interaction between the librations and the bending band (red dashed line) is provided. Spectral functions obtained through the interaction of bending band with themselves (green dot dashed line) and librations with themselves (blue dotted line) are shown too.}
    \label{fig:first_cm}
\end{figure}

We report the results for $\ce{H_2O}$ ice at T = 200K, but the same also holds in deuterated ice. We show in Fig. \ref{fig:first_cm} that the first combination mode, occurring for $\omega \in [2250,2800] \si{\centi\metre^{-1}}$, is originated from the interaction between the libration and bending bands, as the peak appears only if we account for their reciprocal interaction. If we account only for librational or bending modes in the calculations, the peak vanishes. The residual mismatch in the low-frequency tail with the full spectral function reveals a nonnegligible contribution of the other phonon branches (mainly translations).

\begin{figure}
    \centering
    \includegraphics[width=8.6cm]{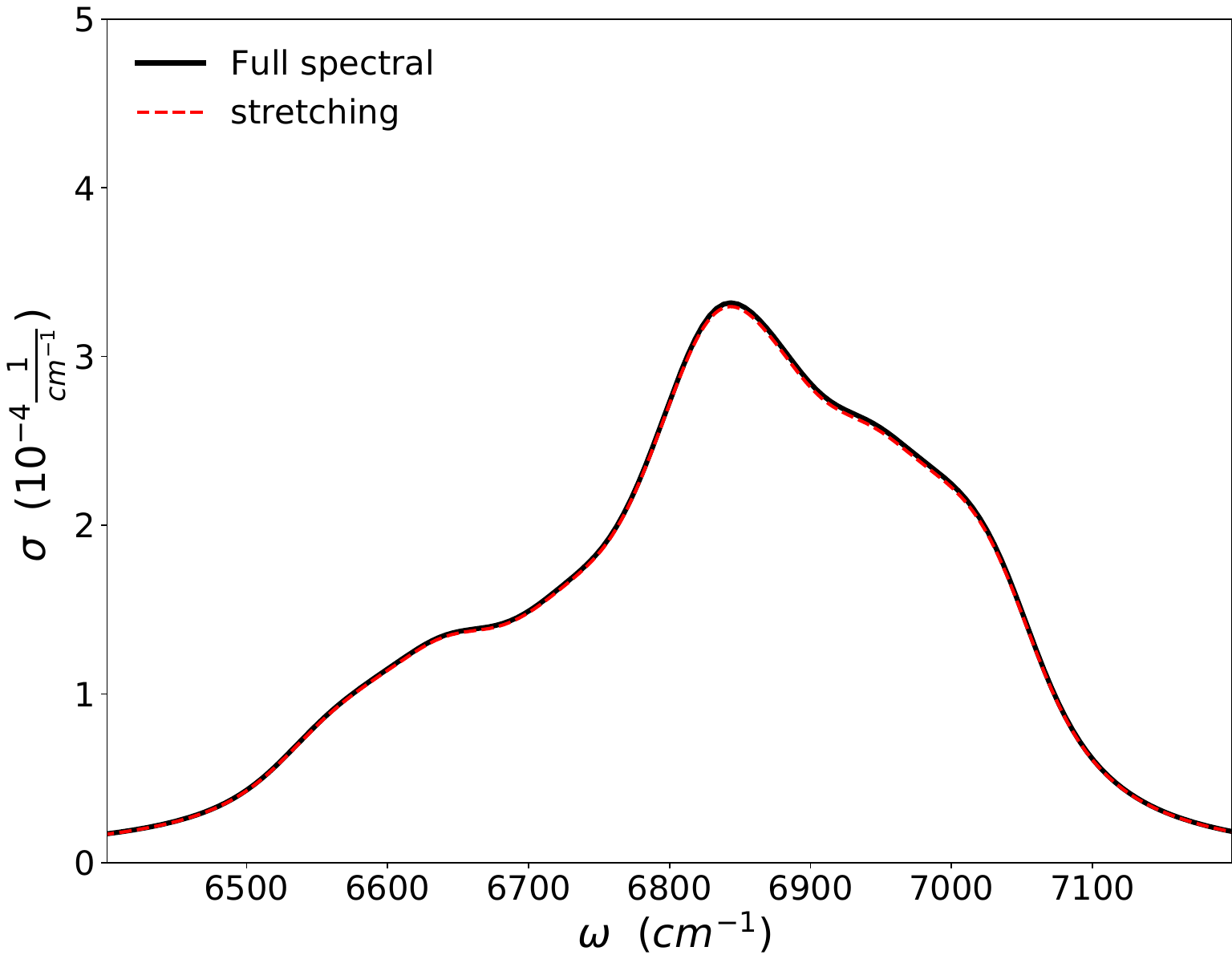}
    \caption{Spectral function of $\ce{H_2O}$ ice at T = 200 $\si{K}$ in the energy range of the second combination mode at $\Gamma$. The full spectral function (solid black line) is compared to that obtained by selecting only the interaction of the stretching band with itself (red dashed line). }
    \label{fig:second_cm}
\end{figure}

The overtone at twice the stretching frequency is analyzed in Fig. \ref{fig:second_cm}. The perfect matching between the full spectral function and the same computed only considering stretching modes unveils how this peak is entirely generated by stretching modes interacting with themselves, without a significative contribution of other phonon branches. It is, in fact, the overtone of the stretching modes.

From the spectral function, we can extract the real phonon energy and their lifetimes, as shown in App. \ref{app:methods}. We report in Table \ref{tab:linewdith} the phonon energies and their linewidth for few selected intense modes at T = 0 K and T = 300 K in the Brillouin zone center.

\begin{table}
\centering
\subfloat[T=0 $\si{K}$]{
\begin{tabular}{ccc}
& $\bbOmega_\mu$ [$\si{\centi\metre^{-1}}$] & $\Gamma_\mu$ $[\si{\centi\metre^{-1}}]$\\
\hline
& 208  & 12\\
tr. &326  & 11\\
\hline
&602  & 4\\
&740  & 5\\
lib. &851  & 12\\ 
&968  & 16\\
\hline
&1565 & 16\\
bend. &1606 & 20\\
\hline
&3102 & 13\\
str. &3207 & 16\\
&3399 & 47\\
\end{tabular}}
\qquad
\subfloat[T=300 $\si{K}$]{
\begin{tabular}{cc} 
$\bbOmega_\mu$ [$\si{\centi\metre^{-1}}$] & $\Gamma_\mu$ $[\si{\centi\metre^{-1}}]$\\
\hline
211  & 75\\
311  & 52\\
\hline
557  & 20\\
682  & 43\\
808  & 47\\
940  & 54\\
\hline
1551 & 31\\
1594 & 40\\
\hline
3080 & 68\\
3237 & 77\\
3439 & 97\\
\end{tabular}}
\caption{Peak frequency and linewidths of some selected intense modes in $\ce{H2O}$ ice at ambient pressure for two values of temperature, T=0 $\si{K}$ and 300 $\si{K}$ at $\Gamma$. The first column indicates to which band the modes belong. See App. \ref{app:methods} for details about the calculations of frequencies and linewidths.}
\label{tab:linewdith}
\end{table}

We observe an essential reduction of the lifetime increasing the temperature. The phonon energies are less temperature-dependent than their lifetime. Stretching modes gain energy upon heating while all the others become softer.

\subsection{Spectroscopy}
\label{sec:Dynamical_results}

Anharmonicity shifts the frequencies of phonons and introduces a finite lifetime.  Here, we simulate the Raman and IR vibrational spectroscopy on ice, adequately accounting for quantum and thermal anharmonic nuclear motion.

Raman scattering and infrared absorption are complementary tools to probe phonon energies at $\Gamma$. The former is based on an inelastic scattering process detecting modes due to changes in the polarizability, while the latter relies on the absorption process and the vibrations detected involve modifications of the dipole moment. Consequently, the selection rules for the two spectroscopies are different, and often active IR modes are Raman inactive (or vice versa).

Appendix \ref{app:Raman_IR} and ref.\cite{Monacelli2020,Monacelli2021} describe the relationships between Raman and IR signal and the anharmonic phonon Green functions.

\begin{figure}[hbtp]
    \centering
    \includegraphics[width=8.4cm]{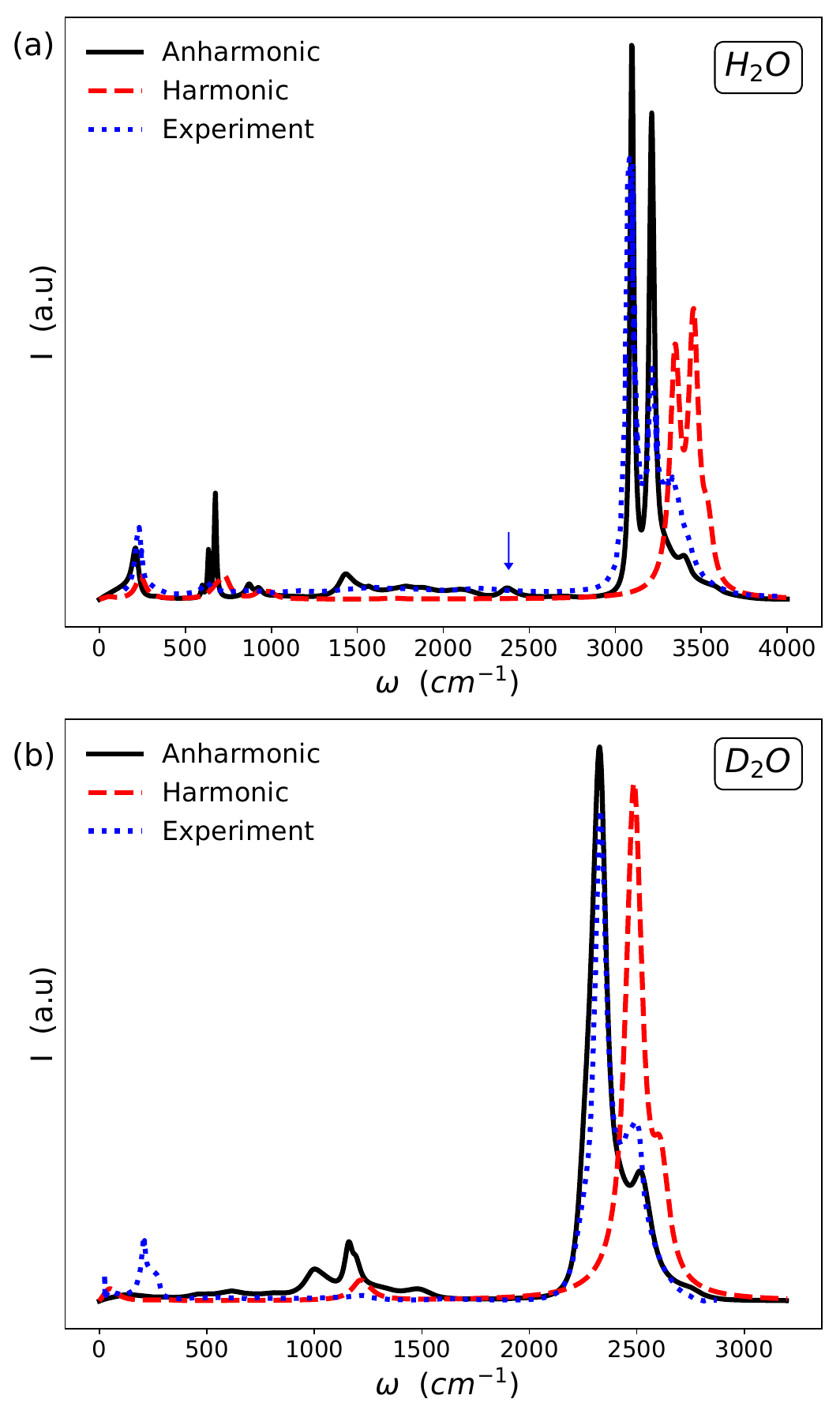}
    \caption{$\mathbf{a}$ Raman scattering spectra at T=65K in $\ce{H_2O}$ ice XI in the a(c,*)b geometry. The experimental spectrum \cite{Shigenari2012}  (blue dotted line) is compared with the harmonic (red dashed line) and the anharmonic phonons (solid black line), computed in the SSCHA framework with the inclusion of the bubble term in Eq. (\ref{eq:self_bubble}). $\mathbf{b}$ Raman scattering spectra for deuterated ice at T=269K in the c(a,a)b geometry. Comparison between harmonic (red dashed line), anharmonic SSCHA phonons with the bubble correction (solid black line) spectra, and the experiment \cite{Scherer1977} (blue dotted line). Blue arrows indicate the position of the combination mode. An artificial broadening is employed in the harmonic approximation to guide the eyes in the comparison of the spectrum with experiment of \SI{35}{\per\centi\meter} ( \SI{45}{\per\centi\meter}) in $\ce{D_2O}$ ($\ce{H_2O}$). Instead, the broadening of the anharmonic simulation is fully obtained \emph{ab initio} from phonon-phonon scattering.}
    \label{fig:Raman_scattering}
\end{figure}

In Fig. \ref{fig:Raman_scattering} ($\mathbf{a}$) , we compare the simulated Raman spectra with the experiment \cite{Shigenari2012} for the same geometry c(a,*)b (see App. \ref{app:Raman_IR}) in $\ce{H_2O}$ ice XI at T=65 K. We report deuterated ice at T=269K in the c(a,a)b geometry \cite{Scherer1977} in Fig. \ref{fig:Raman_scattering} ($\mathbf{b}$) .

The result obtained in the anharmonic dynamical theory (TD-SSCHA) matches perfectly with the experimental results, correcting a shift of the harmonic phonon energy in the stretching modes of about $10\%$ (7$\%$) of the energy in $\ce{H_2O}$ ($\ce{D_2O}$) ice. The theory can predict the presence of the combination mode (indicated by the blue arrow in Fig. \ref{fig:Raman_scattering} ($\mathbf{a}$) ) for $\ce{H_2O}$ ice in the considered geometry. Instead, this mode has very low intensity in the experimental geometry of $\ce{D_2O}$ ice.

Finite linewidths in the harmonic model are for presentation purposes only, as harmonic phonons have infinite lifetimes.

\begin{figure}[hbtp]
    \centering
    \includegraphics[width=\columnwidth]{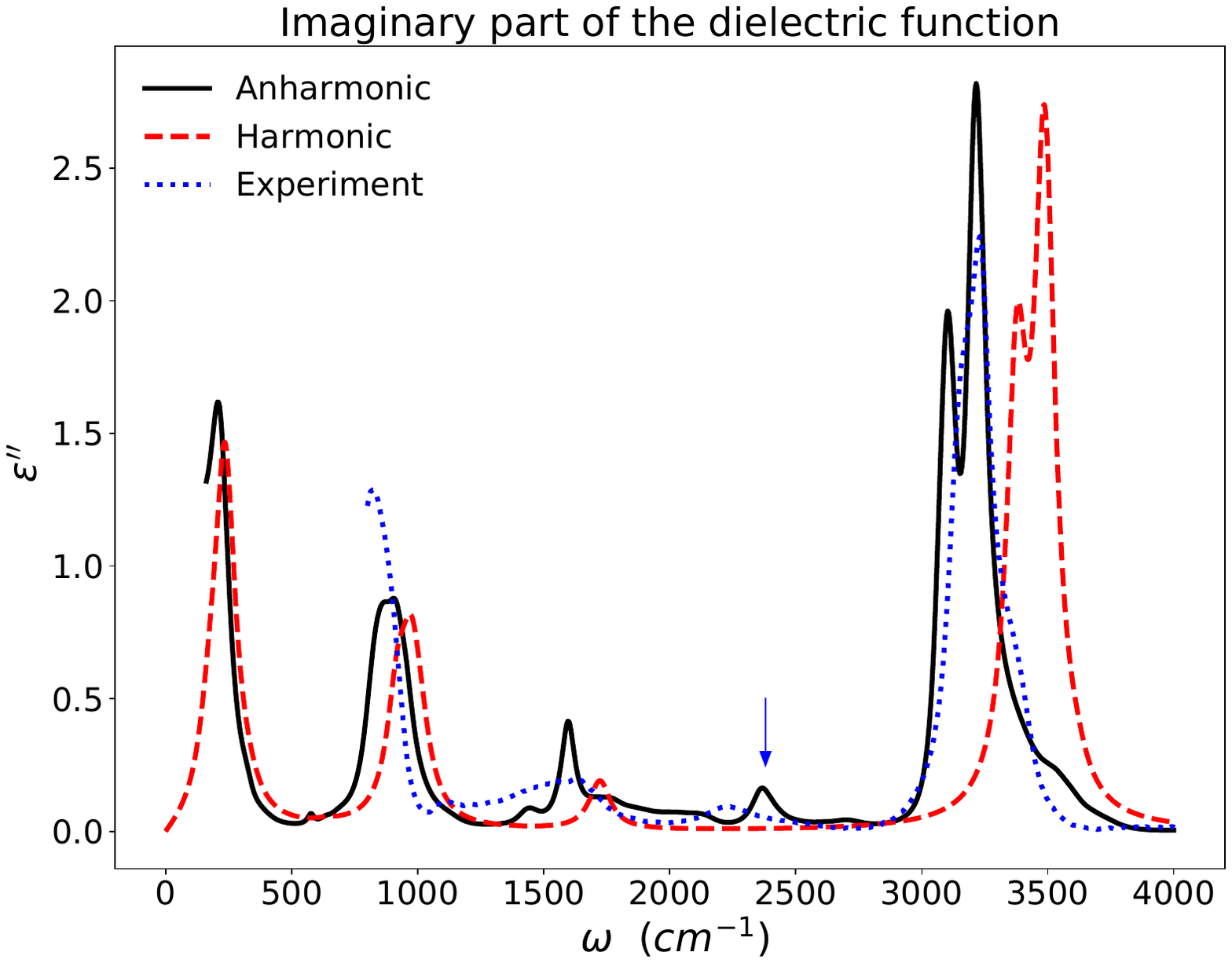}
    \caption{Imaginary part of the dielectric function of $\ce{H_2O}$ ice at T=200K.
    The trace of the dielectric tensor as in Eq. (\ref{eq:trace_dielectric}) is reported to take into consideration all the possible orientation of the crystal. The results within the harmonic approximation (red dashed curve) and the anharmonic phonons (black solid line), computed in SSCHA framework with the inclusion of the bubble term of Eq. (\ref{eq:self_bubble}), are compared with the experiment \cite{Clapp1995} (blue dotted line). The blue arrow indicates a combination mode. We employed an artificial broadening of \SI{50}{\per\centi\meter} in the harmonic spectrum. Instead, the broadening of the anharmonic simulation is fully obtained \emph{ab initio} from phonon-phonon scattering.}
    \label{fig:Dielectric}
\end{figure}

Phonon vibrations impact the low energy optical properties of any material. In Fig. \ref{fig:Dielectric}, we report the imaginary part of the dielectric function at \SI{200}{\kelvin} of $\ce{H_2O}$ ice. To include all the crystal orientations, we plot the trace of the dielectric tensor as in Eq. (\ref{eq:trace_dielectric}).  As for the Raman, anharmonicity reduces by $10\%$ the harmonic stretching band energy, providing perfect agreement with the experiment \cite{Clapp1995}. Also, the combination mode located at \SI{2300}{\per\centi\meter} (highlighted by the blue arrow in Fig. \ref{fig:Dielectric}) observed experimentally is correctly reproduced by the anharmonic spectrum.

Both the dielectric function of Fig. \ref{fig:Dielectric} and the Raman scattering spectra in Fig. \ref{fig:Raman_scattering} confirm the importance of anharmonicity. They further demonstrate how it is fundamental to reproduce the experimental results and provide the SSCHA (and its time-dependent extension) as the best tool for ice simulation.

\section{Conclusions}
\label{sec:conclusion}

Exploring the thermodynamic structural and vibrational properties of ice XI (hydrogen-ordered counterpart of ordinary ice $I_{\textit{h}}$), we unveiled the importance of quantum anharmonic effects. The anomalous strong temperature dependence of the bulk modulus, $20\%$ variation from 0 to 300K, is fully explained by thermal and quantum anharmonic fluctuations, revealing the combined effect of the vibrations ($64\%) $ and the thermal expansion (36$\%$).

We unmasked the failure of the state-of-art techniques in reproducing the anomalous VIE. Only a complete anharmonic treatment of quantum nuclear motion enables the reproduction of the experimental results. In particular, we proved how the negative VIE originates from a nonmonotonous volume expansion compared with quantum fluctuations. If we increase the mass of the hydrogen isotopes, the volume first expands, saturating slightly above the tritium mass, and then contracts to the classical value. This means that the VIE is due to a strongly nonlinear regime of quantum fluctuations in ice, which commonly employed state-of-the-art theories (as the QHA) do not grasp.
Notably, also oxygen is in a strong quantum mechanical regime, being responsible for a $2\%$ volume reduction in the classical limit. 

We observe an anharmonic renormalization of $8-10\%$ in the bending and stretching modes that grants a good prediction of Raman scattering spectra and dielectric function of $\ce{H_2O}$ and $\ce{D_2O}$ ice XI.

For the first time, the low-energy range of phonon dispersion of deuterated ice at T=140K is excellently reproduced by the anharmonic renormalized phonons and correct treatment of the electron exchange and correlation. This result paves the way for the study of thermal transport from first principles and the simulation of ice under pressure, where acoustic phonons are the only modes detectable.

Our simulations deciphered the microscopic origin of many anomalous properties of ice, proving how anharmonicity and quantum fluctuations of ions are a mandatory ingredient to reproduce the thermodynamic structural and vibrational properties of ice.

\begin{acknowledgements}
The authors acknowledge the CINECA award under the ISCRA initiative, for the availability of high performance computing resources and support.
\end{acknowledgements}

\section*{Data Availability Statement}
The data that support the findings of this study are available upon reasonable request from the authors.

\appendix
\section{Theoretical Methods}
\label{app:methods}

In Sec. \ref{subsec:dispersion_results}, we mentioned that the SSCHA dynamical matrix is not suited to describe real phonons \cite{Bianco2017,Monacelli2021}. This happens because it is positive-definite by construction so that, for example, phonons instabilities cannot occur. The eigenvalues of the Hessian matrix of Eq. (\ref{eq:hessian}) are the response to a static external perturbation and describe the stability of the structure with respect to a spontaneous symmetry breaking.

\begin{equation}
    \label{eq:hessian}
    \bDF= \frac{1}{\sqrt{\mathbf{M}}} \singledot \pdv{\mathcal{F}}{\rscha}{\rscha} \singledot \frac{1}{\sqrt{\mathbf{M}}}
\end{equation}

Here, $\mathcal{F}$ is a short-hand notation for the SSCHA free energy in Eq. (\ref{eq:SSCHA_free_energy}). The Hessian matrix can be written in term of the third- and fourth-order force constant matrices,  $\bPhithree,\bPhifour$, as in \cite{Bianco2017,Monacelli2021}, where,

$$ \overset{\sss{(n)}}{\bPhi} = \expval{ \frac{\partial^n \mathcal{F}} {   \underbrace{ \partial \rscha ...  \partial \rscha }_{\text{n}} } }_{\rhoschatrial} $$

In this work, we use the lowest order correction in the Hessian matrix, the one containing the third-order force constant matrix, after checking that the fourth-order contribution is negligible. In these conditions, the free energy Hessian can be approximated as in Eq. (\ref{eq:Hessian_expansion}),

\begin{equation}
    \label{eq:Hessian_expansion}
    \bDF \simeq \bDscha + \bDthree \doubledot \bLambda [0]  \doubledot \bDthree
\end{equation}
where $\bDscha$ is the SSCHA dynamical matrix and the other term is called the bubble correction (as, at lowest order perturbation theory, it give rise to the bubble diagram)  \cite{Bianco2017}.
$\overset{\sss{(n)}}{\bD}$ is the n-th order force constant matrix divided by the square root of the masses, (the superscript ${}^{(2)}$ is neglected for the second order SSCHA dynamical matrix)

\begin{equation}
    \label{eq:D_matrix}
    \overset{\sss{(n)}}{\bD} = \frac{\overset{\sss{(n)}}{\bPhi}}{\underbrace{ \sqrt{\bM}...\sqrt{\bM}}_{\text{n}}} 
\end{equation}

$\bLambda[0]$ is the zero frequency value of the fourth-order tensor that depends only on the eigenvalues $\omega_\mu$ and eigenvectors $\bm\epsilon_\mu$ of the SCHA auxiliary dynamical matrix $\bDscha$.

\begin{equation}
    \label{eq:Lambda_tensor}
    (\Lambda[z])^{abcd} = \sum_{\mu \nu=1}^{3N} \mathscr{F}(z,\omega_\mu,\omega_\nu) \epsilon_\nu^a \epsilon_\mu^b \epsilon_\nu^c \epsilon_\mu^d
\end{equation}

\begin{align}
    \label{eq:f_function}
    \mathscr{F}(z,\omega_\mu,\omega_\nu) = -& \frac{1}{4\omega_\mu\omega_\nu}
    \frac{(\omega_\mu + \omega_\nu)(n_\mu + n_\nu + 1)}{(\omega_\mu + \omega_\nu)^2 - z^2} + \nonumber \\
    & \frac{1}{4\omega_\mu\omega_\nu}
    \frac{(\omega_\mu - \omega_\nu)(n_\mu - n_\nu) }{(\omega_\mu - \omega_\nu)^2 - z^2}
\end{align}

The $a,b,c,d $ indices run over the atoms in the supercell and the Cartesian coordinates.

The $\bLambda$ tensor describe the propagation of the $\mu$ $\nu$ phonon modes, whose interaction can give rise to combination modes, as seen in Fig. \ref{fig:dos_combination_modes}. By restricting the sum in Eq. \ref{eq:Lambda_tensor}, it is possible to isolate the contribution of few selected modes to the free energy Hessian.

The eigenvalues of the Hessian matrix can be used to approximate the real phonons in the low energy regime, close to z=0, as in Fig. \ref{fig:Dispersion_exp} ($\mathbf{a}$). 
However, physical phonons, those observed by experimental probes like inelastic  scattering and vibrational spectroscopy, must be computed from the dynamical interacting Green function.
Within the SSCHA framework, the dynamical Green function  $\bG(z)$ for the displacement normalized to the masses, $\sqrt{\bM}(\bR - \rscha)$, in component free notation, is \cite{Bianco2017,Monacelli2021}

\begin{equation}
    \label{eq:Green_componentfree}
    \bG^{-1}(z) = z^2 \mathbb{1} -( \bDscha - \bPi(z))
\end{equation}

$\bDscha$ is the SSCHA dynamical matrix at equilibrium. The full expression for the SSCHA self-energy $\bPi(z)$ can be found in \cite{Bianco2017,Monacelli2021}.

In this work, as for the free energy Hessian, we decide to keep the lowest order of the self-energy correction, the bubble $\bPiB (z)$,

\begin{equation}
    \label{eq:self_bubble}
    \bPi(z) \simeq \bPiB(z) = \bDthree \doubledot \bLambda (z) \doubledot \bDthree
\end{equation}

The real phonons associated are the poles of the dynamical Green function (See \cite{Monacelli2021}, Sec. IV C, for further details on the computation of the poles $\bbOmega_\mu$ and linewidths $\Gamma_\mu$).

\section{Computational details}
\label{app:compuational_details}
To simulate the thermodynamic properties, we computed the harmonic free energy (relaxing the atomic position at fixed cell) at 120 volumes. The free energy is fitted with the Vinet equation of state (EOS) \cite{Vinet1987} to obtain the equilibrium volume and the bulk modulus for any temperature as in Eq. (\ref{eq:equilibrium_volume}) and Eq. (\ref{eq:bulk_modulus}). 

In contrast, for the SSCHA, we computed the pressure as a function of temperature for six volumes. At each volume, we relaxed the atomic positions accounting for quantum and thermal anharmonic effects. We employed ensembles with as many as 100000 configurations in the converged supercell to reduce the statistical noise. We evaluated the equilibrium volume and the bulk modulus fitting the $P(\Omega)$ curve with the Vinet EOS at each temperature.

For QHA simulations, we sampled phonons in a 14x14x14 mesh of the Brillouin zone; for the SSCHA ones, we employed a 3x3x2 supercell. 
The convergence tests are reported in appendix~\ref{app:convergence}.

The spectral properties are computed from the dynamical one-phonon interacting Green function. The self-energy is approximated as in Eq. \ref{eq:self_bubble} and computed integrating on 14x14x14 a $\bk$-grid in the reciprocal space \cite{Bianco2018}. Furthermore, a smearing factor $\delta_{se}$ is introduced to obtain converged results in the computation.

\begin{equation}
    \label{eq:smearing}
    \bPi(z) \simeq \bPiB(z) = \bPiB(z+i \delta_{se})
\end{equation}

Convergence is achieved for $\delta_{se} = 45 \si{\centi\metre^{-1}}$. The same value holds for $\ce{H_2O}$ and $\ce{D_2O}$ ice. Consequently, all the SSCHA spectral functions are computed with those values of smearing and integration $\bk$-grid. Finite linewidths in the DOS and harmonic model are for presentation purposes only.

\section{Raman and Infrared}
\label{app:Raman_IR}

The Raman spectrum is proportional to the polarizability correlation function $\expval{\alpha(t)\alpha(0)}$, where

\begin{equation}
    \label{eq:polarizability}
    \alpha_{ab}(t)= \sum_{c=1}^{3N} A_{abc}u_c(t)
\end{equation}

The Raman tensor $A_{abc}$ is computed \emph{ab initio} in LDA approximation. Here, a and b are Cartesian indices and c is a super-index running over the Cartesian coordinates and the atoms in the supercell. We keep this notation in the following.

The intensity of the Raman signal has been evaluated as:
\begin{equation}
    \label{eq:Raman_intensities}
    I^{\text{Raman}}(\omega) \propto (\omega -\omega_L)^4 \sum_{a,b=1}^{3N} \frac{A'_aA'_b}{\sqrt{M_aM_b}} \Im G_{ab}(\omega)
\end{equation}
where 
$\omega_L$ is the frequency of the laser and $A'$ accounts for the polarization of the incident ($\epsilon^{\text{in}}$) and scattered ($\epsilon^{\text{out}}$) light \cite{Shigenari2012,Abe2011,Scherer1977}:
$$ A'= \mathbf{\epsilon^{\text{in}}} \singledot A \singledot \mathbf{\epsilon^{\text{out}}} $$

In most cases, the laser frequency is much bigger than the phonon frequencies, so that the approximation $|\omega-\omega_L |\simeq \omega_L$ holds. This is not completely true in ice, where phonons can be very energetic ( $\omega \simeq 3400 $  $\text{cm}^{-1}$). Here, that difference provides a slight reshaping of the spectra, and it is safer to avoid approximations.

\begin{figure}
    \centering
    \includegraphics[width=\columnwidth]{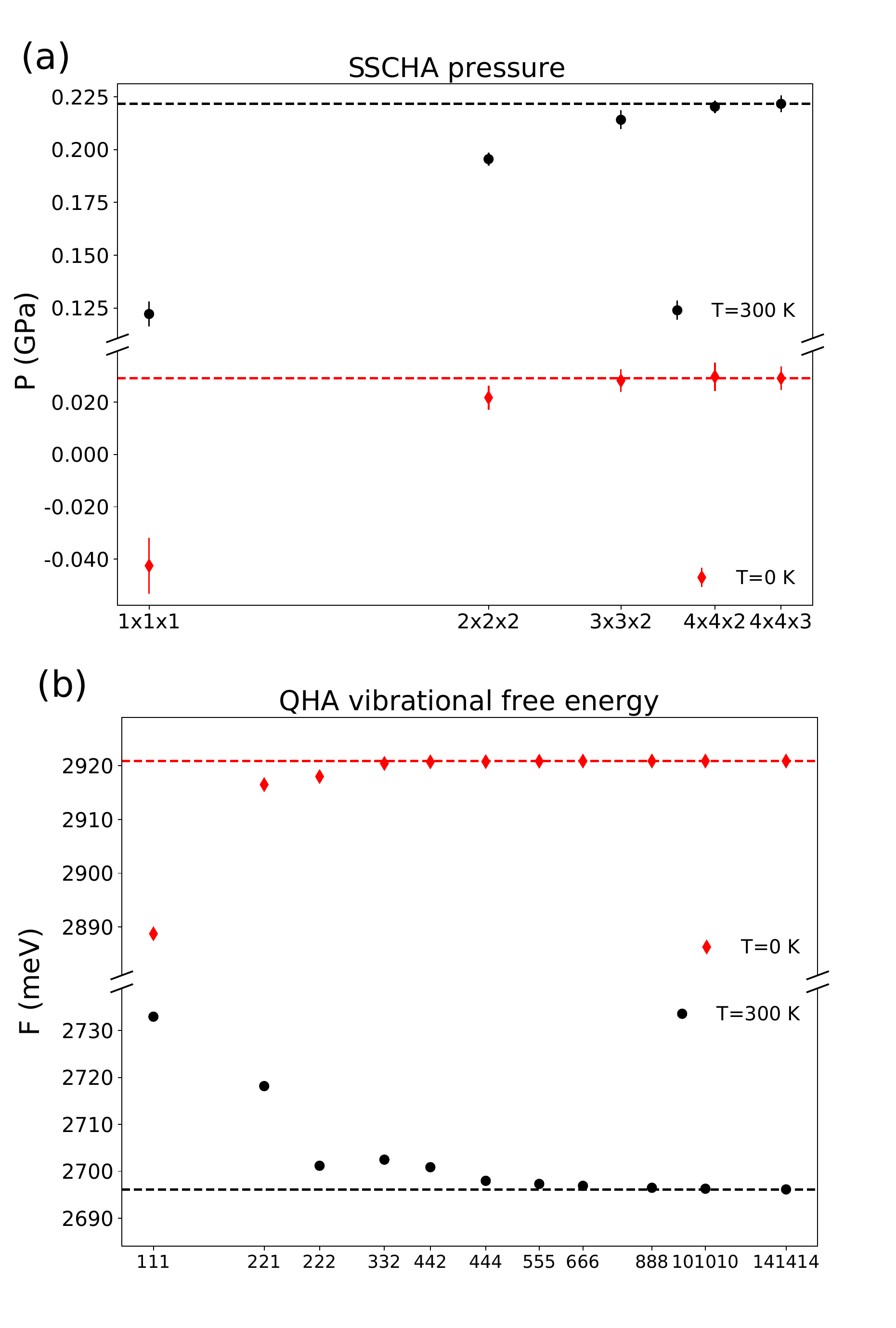}
    \caption{Convergence tests in the QHA and SSCHA. $\mathbf{a:}$ Pressure within the SSCHA framework computed as in Ref. \cite{Monacelli2018} as a function of the supercell dimension for T=0 K and T = 300K $\mathbf{b:}$ Vibrational term of the QHA free energy in Eq. (\ref{eq:QHA_definition}) as a function of the supercell dimension for  T=0 K and T = 300K. Here we use the short hard  notation $\textit{N}_1 \textit{N}_2 \textit{N}_3$ to indicate the $ \textit{N}_1 \times \textit{N}_2 \times \textit{N}_3$ supercell. The dashed lines in panel  $\mathbf{a}$ and $\mathbf{b}$ are the converged results for the pressure or the free energy at the given temperature.}
    \label{fig:Convergence}
\end{figure}

The physical quantity determining the Infrared absorption is the dielectric function and standard electromagnetism provides a simple relation between the dielectric tensor and the susceptibility.

\begin{equation}
    \label{eq:susceptibility}
    \vb{\epsilon}(\omega)= 1 + 4\pi \vb{\chi}^{(\text{tot})} (\omega)= \vb{\epsilon}^{\text{el}} + 4\pi \vb{\chi}^{\text{ion}}(\omega)
\end{equation}

The electronic part $\epsilon^{\text{el}}$ is computed \emph{ab initio}. The absence of electronic transitions in the phonon energy range makes it real and it is frequency independent, $\epsilon^{\text{el}}=1.65$.
The ionic susceptibility of Eq. (\ref{eq:susceptibility}) is the Fourier transform of the dipole-dipole correlation function:

\begin{equation}
    \label{eq:susceptibility_IR}
    \chi^{\text{ion}}_{ab}(\omega) = \int dt e^{-i \omega t} \expval{ M_a(t)M_b(0)}
\end{equation}

where, $M_a(t)= |e| \sum_{b=1}^{3N} Z_{ab} u_b(t)$, the effective charges $Z$ are computed \emph{ab initio} in Quantum Espresso \cite{Giannozzi_2009} in LDA approximation \textcolor{blue}.

We computed the dielectric function as:
\begin{equation}
    \label{eq:dielectric_final}
    \epsilon_{\alpha \beta}(\omega)= \epsilon^{\text{el}}_{\alpha \beta} + 4\pi |e|^2 \sum_{ab} \frac{Z_{\alpha a}Z_{\beta b}}{\sqrt{M_aM_b}} G_{ab}(\omega)
\end{equation}
and averaged over all possible orientation of the crystal:
\begin{equation}
    \label{eq:trace_dielectric}
    \epsilon(\omega)= \frac{1}{3} \sum_{\alpha=1}^3 \epsilon_{\alpha \alpha}(\omega)
\end{equation} 

\section{Convergence}
\label{app:convergence}

\begin{figure}
    \centering
    \includegraphics[width=\columnwidth]{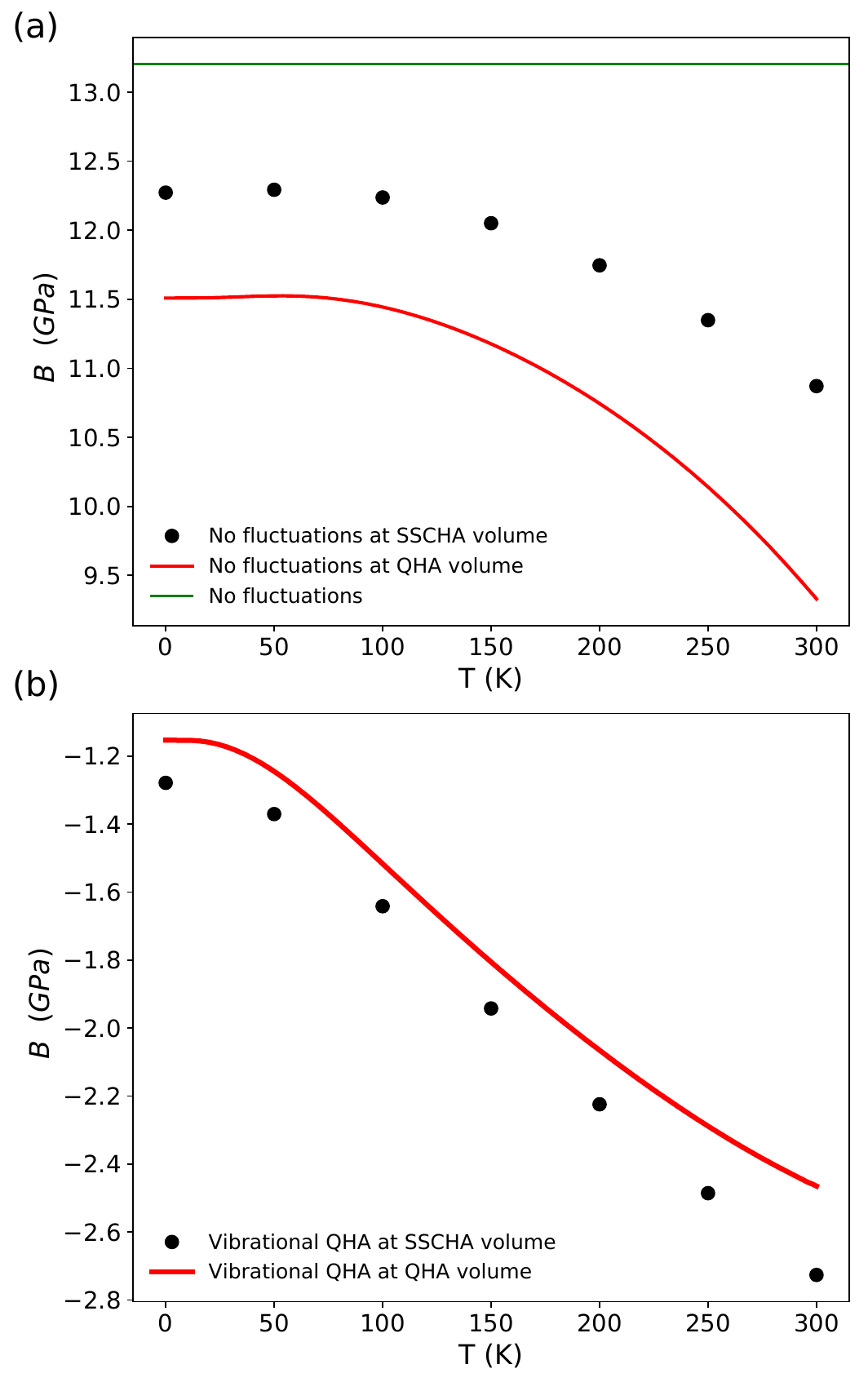}
    \caption{Analysis of the different contributions to the bulk modulus. $\mathbf{a}$ The static QHA bulk modulus is computed at the QHA equilibrium volumes (solid red line) and  at the SSCHA equilibrium volumes (black circles). The green solid line is the classical bulk modulus where fluctuations are neglected. $\mathbf{b}$ Vibrational QHA bulk modulus computed at the SSCHA (black circles) and QHA (red solid line) equilibrium volumes.}
    \label{fig:bulk_analysis}
\end{figure}

Eq. (\ref{eq:QHA_definition}) and Eq. (\ref{eq:SSCHA_free_energy}) depend on the number $N_q$ of q points in the Brillouin zone.
We employ the supercell method, that  consists in replicating an integer number of times $ \textit{N}_1 \times \textit{N}_2 \times \textit{N}_3$ the unit cell along the three Cartesian directions and imposing periodic boundary conditions.
The unit cell of ice XI has an orthorhombic structure with $Cmc2_1$ space group and contains 4 water molecules.
In Sec. \ref{subsec:Thermodynamic_results}, we pointed out that the thermodynamic properties in the QHA are derived from the free energy, while in the SSCHA the pressure has been used, thanks to the availability of a handy analytical formula \cite{Monacelli2018}.

Consequently, the converged supercells are chosen by looking at the free energy in the QHA and at the pressure in the SSCHA. We analyzed the convergence for the two extremal temperatures, T=0K and T=300K, in Fig. \ref{fig:Convergence}. As a general trend, we notice that thermal fluctuations slow down the convergence of both the free energy and the pressure. The limited computational cost of the QHA doesn't place any limitation on the mesh to use. 

Moreover, we have the possibility to interpolate the dynamical matrix to a finer mesh, as it has been done for the two bigger meshes, namely the $ 10 \times 10 \times 10$ and the $14 \times 14 \times 14$.
So, by looking at Fig. \ref{fig:Convergence} ($\mathbf{b}$) we decided to use the biggest grid we analyzed.

Conversely, the challenging computational cost of the SSCHA prevents the use of a big supercell. Fig. \ref{fig:Convergence} ($\mathbf{a}$) reveals as the $3\times3\times2$ supercell grants at most 3$\%$ error with respect to the converged mesh at high temperature. We are satisfied with this accuracy level. We compute the thermodynamic properties in this mesh. Instead, for the phonon dispersion of Sec. \ref{subsec:dispersion_results}, a 4x4x4 supercell has been used.

\section{Bulk modulus}
\label{app:bulk_modulus}

The bulk modulus is computed from Eq. (\ref{eq:bulk_modulus}). The free energy consists in a static and a vibrational term as in Eq. (\ref{eq:QHA_definition}), so, the same is for the bulk modulus. In the QHA picture, the static and the vibrational part are in Eq. (\ref{eq:bulk_separation}):

\begin{subequations}
\label{eq:bulk_separation}
\begin{align}
    B_{\textit{stat}} (T) & =  \Omega_{\textit{eq}}(T) \pdv[2]{V(\rscha, \lbrace \vec{a}_i \rbrace)}{\Omega} \bigg |_{\Omega_{\textit{eq}(T)} } \label{eq:bulk_stat} \\
    B_{\textit{vib}} (T) & = \Omega_{\textit{eq}}(T) \pdv[2]{F_{\textit{vib}}(\rscha, \lbrace \vec{a}_i \rbrace)}{\Omega} \bigg |_{\Omega_{\textit{eq}(T)} } \label{eq:bulk_vib} 
\end{align}
\end{subequations}

The division of the bulk modulus into its two contributions makes possible to individuate the origin of its strong temperature dependence observed in Fig. \ref{fig:bulk}.

Fig. \ref{fig:bulk_analysis} ($\mathbf{a}$) shows the static bulk modulus, where fluctuations are neglected. We computed the curves at the QHA and SSCHA equilibrium volumes, in order to introduce the effect of thermal expansion. The continuous line is the classic result, where quantum and thermal effects are not included.
The vibrational contribution of Eq. (\ref{eq:bulk_vib}), for the QHA and SSCHA volumes, is reported in Fig. \ref{fig:bulk_analysis} ($\mathbf{b}$). It is always negative, meaning that it would increase the volume under compression.

Both the static and the vibrational bulk modulus have a non negligible temperature dependence, contributing to the total one for the $64 \%$ and $36\%$ respectively.

Moreover, the effect of different equilibrium volumes is almost temperature independent in the vibrational term, being unable to explain the different thermal behaviour of the two theories, that can instead be addressed partially to the effect of volumes in the static bulk modulus as evident in Fig. \ref{fig:bulk_analysis} ($\mathbf{a}$).

\section{Dispersion}
\label{app:dispersion}

\begin{figure}
    \centering
    \includegraphics[width=\columnwidth]{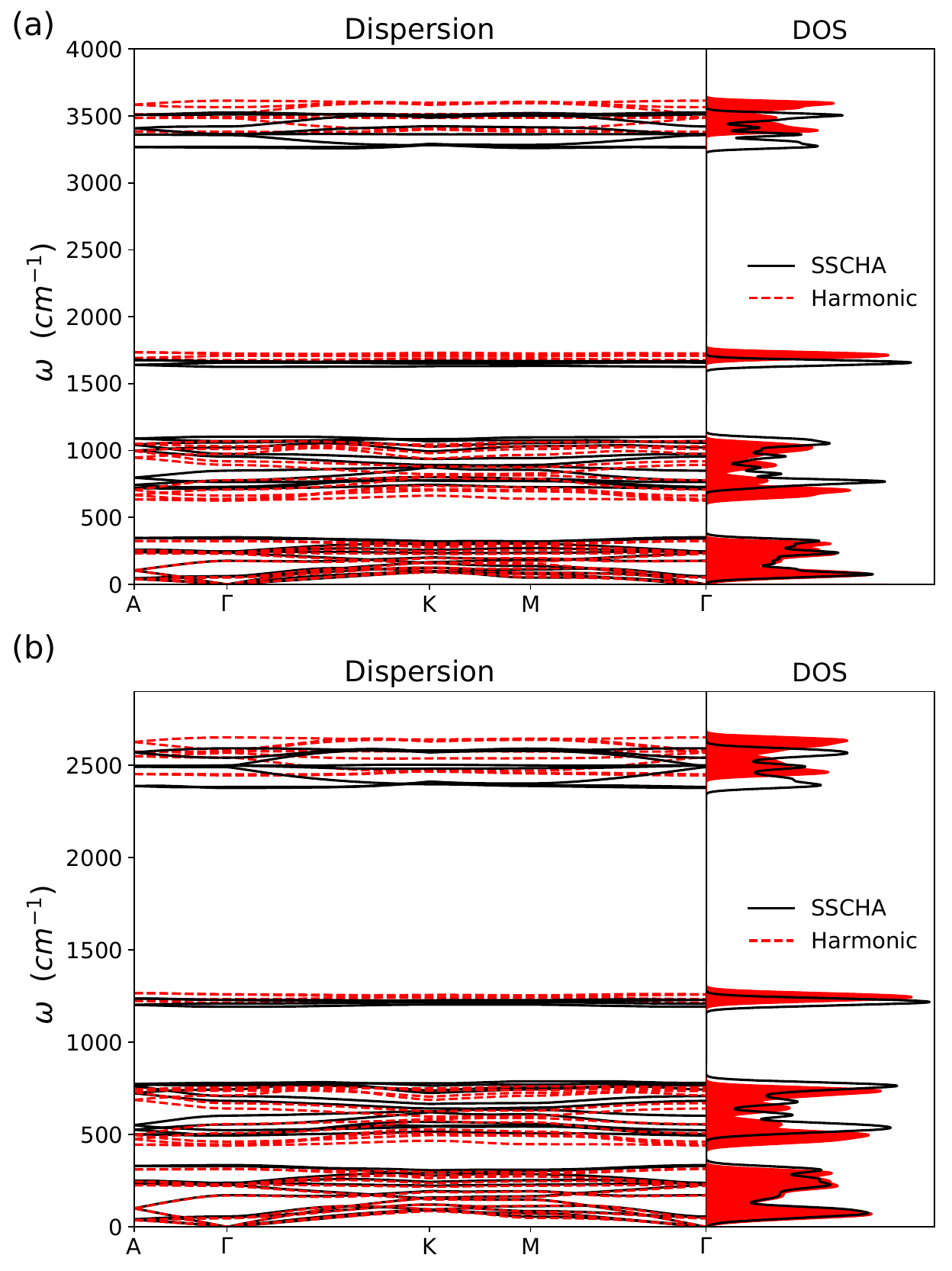}
    \caption{Phonon dispersion and density of states.  $\mathbf{a}$  $\ce{H_2O}$ ice XI at T=200K at ambient pressure$. \mathbf{b}$ $\ce{D_2O}$ ice XI at T=140K and P=0.05 GPa. In both panel $\mathbf{a}$ and $\mathbf{b}$, the harmonic results (red dashed line) are compared with the dispersion and DOS in the SSCHA framework (black solid lines).}
    \label{fig:DOS_dispersion}
\end{figure}

\begin{figure}
    \centering
    \includegraphics[width=\columnwidth]{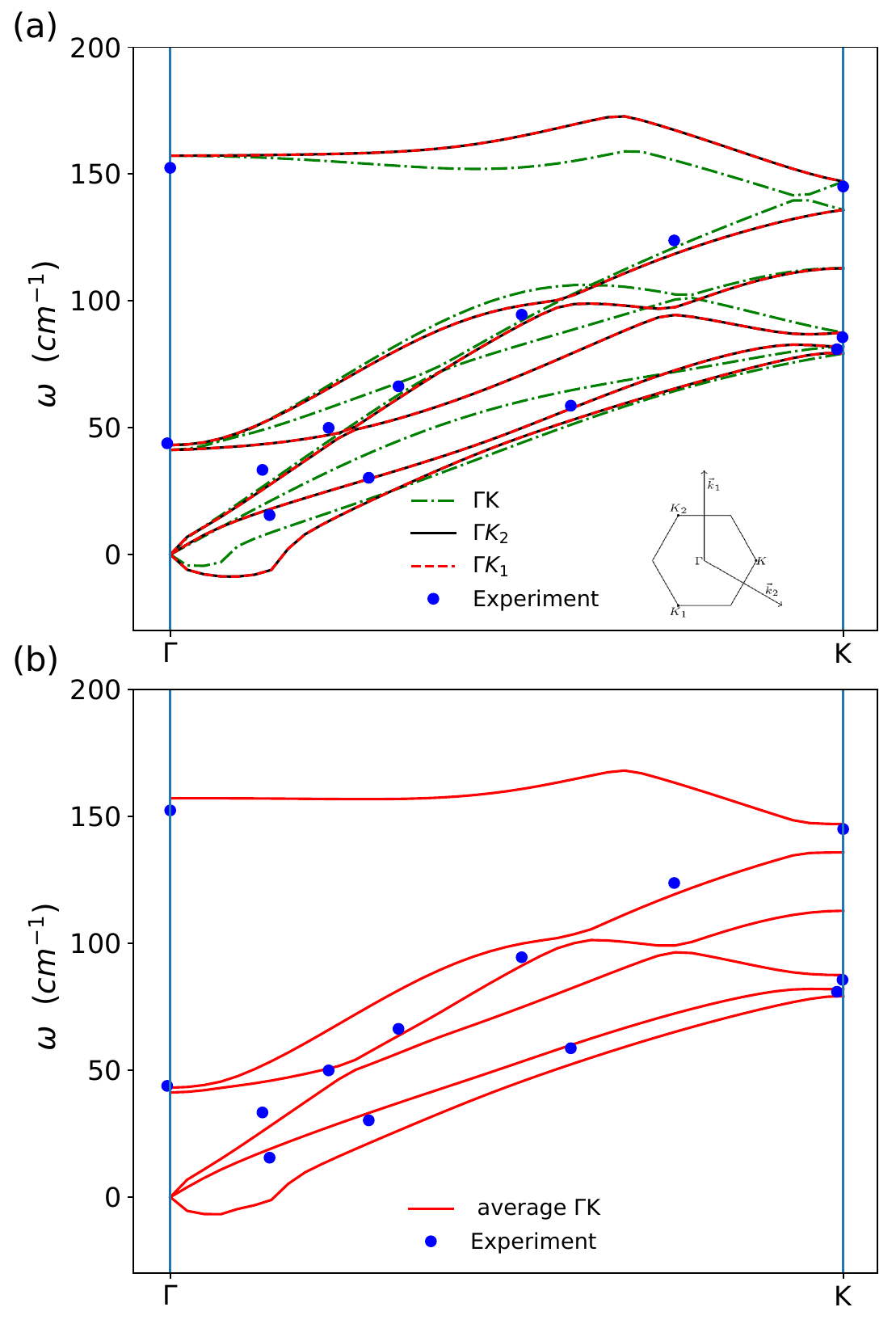}
    \caption{Phonon dispersion in the SSCHA framework with the inclusion of the bubble term in the self-energy correction in the static limit for $\ce{D_2O}$ ice at T=140 K and P = 0.05 GPa. $\mathbf{a}$ Phonon dispersion along the three nonequivalent $\Gamma K$ directions compared to the experimental measurement (blue circles) \cite{Strssle2004}. The three K points and the projection in the plane of the first Brillouin Zone are shown in the inset. $\mathbf{b}$ Comparison of the experiment with the average of the phonon dispersion along the three $\Gamma K$ directions (red solid lines).}
    \label{fig:ineq_dispersion}
\end{figure}

Real phonon dispersion is calculated from the dynamical interacting Green function as in App. \ref{app:methods}.
Here, we show the density of states (DOS) and dispersion for $\ce{H_2O}$ ice at T=200K and $\ce{D_2O}$ ice at T=140K and P=0.05 $\si{GPa}$, comparing the harmonic phonons and the SSCHA auxiliary phonons in Fig. \ref{fig:DOS_dispersion}.

The harmonic dynamical matrix is computed with a 5x5x5 $\bq$ mesh, while for the SSCHA we employed the 4x4x4 supercell.

As a molecular crystal, the phonon branches are well separated in translational modes, librations, narrow bending and stretching from low to high energy. In Fig. \ref{fig:DOS_dispersion} ($\mathbf{a}$), we report the hydrogen and in Fig. \ref{fig:DOS_dispersion}  ($\mathbf{b}$) the deuterium.

In the SSCHA, the harmonic translational and rotational modes suffer a blue shift of the order of $4.2\%$ ($3.5\%$) and $8.2\%$ ($7.5\%$) respectively for $\ce{H_2O}$ ($\ce{D_2O}$) ice. Instead, anharmonicity reduces the frequencies of the most energetic modes. Both the harmonic bending and stretching band are red-shifted of about $3.1\%$ ($2.2-2.5\%$) in $\ce{H_2O}$ ($\ce{D_2O}$) ice.

Acoustic modes play a major role in thermal transport. We compared the simulated phonon dispersion of deuterated ice at T=140K and P=0.05 $\si{GPa}$ in Fig. \ref{fig:Dispersion_exp} ($ \mathbf{b}$) with the experiment. For a correct comparison, real phonons are needed. In the low energy regime, we can use the static approximation of the self-energy \cite{Bianco2017,Monacelli2021}. 
We consider the lowest order self-energy correction, the bubble of Eq.  (\ref{eq:Hessian_expansion}). This approximation is reasonable and routinely employed in many system with hydrogen\cite{Errea2016_2,Paulatto2015}. In the present case, we checked the approximation against the exact static self-energy, where phonon energies change less than \SI{1}{\per\centi\meter}.

While the harmonic phonons can be computed without effort in the supercell described above, the computational cost of the bubble term is huge, and it is not possible to perform the computation in the same supercell used for the SSCHA dynamical matrix $\bDscha$. To overcome this problem, we first computed the bubble correction in a smaller supercell, namely a 3x3x2, and then we interpolated it to a finer supercell to obtain the Hessian matrix.

\begin{equation}
    \label{eq:interpolation_bubble}
    \bDF_{4x4x4} = \bDscha_{4x4x4} + \biggl[ \bDF_{3x3x2} - \bDscha_{3x3x2} \biggr] ^{4x4x4}
\end{equation}

In Fig. \ref{fig:Dispersion_exp} ($\mathbf{a}$), we restricted the original path of the experiment (the same as in Fig. \ref{fig:DOS_dispersion}) to the $\Gamma$-A direction.
We include the $\Gamma$-K path dispersion in Fig.  \ref{fig:ineq_dispersion}. In ice XI, we have 2 inequivalent $\Gamma$-K path originated by the presence of the hydrogen sublattice, that overturns the perfect equivalence we would obtain in the presence of oxygens alone.

The projection of the first Brillouin zone of ice XI in the plane is reported in the inset of Fig. \ref{fig:ineq_dispersion} ($\mathbf{a}$). We show the planar reciprocal vectors, $\lbrace \vec{b}_1, \vec{b}_2 \rbrace$ and the three K points. 
Notably, the three K points are actually equivalent (the phonon frequencies are the same) but the modes along the path connecting $\Gamma$ and K aren't. 

We average the three dispersion and compared the average with the experiment in Fig. \ref{fig:ineq_dispersion} ($\mathbf{b}$). The matching between theory and experiment is very good.


\addcontentsline{toc}{chapter}{bibliografia}
\bibliography{bibliografia}

\end{document}

%% file: commands.tex
\newcommand{\Rscr}{\mathscr{R}}

\newcommand{\Dcal}{\mathcal{D}}
\newcommand{\Pcal}{\mathcal{P}}
\newcommand{\Rcal}{\mathcal{R}}
\newcommand{\Hcal}{{\hat{\mathcal{H}}}}
\newcommand{\Fcal}{\mathcal{F}}
\newcommand{\tFcal}{{\tilde {\Fcal}}}
\newcommand{\Zcal}{\mathcal{Z}}
\newcommand{\Wcal}{\mathcal{W}}

\newcommand{\ft}{\text{f}}

\newcommand{\bbf}{\mathbbm{f}}
\newcommand{\bbV}{\mathbb{V}}
\newcommand{\Vcal}{{\mathcal V}}
\newcommand{\bnot}{\boldsymbol{0}}
\newcommand{\bA}{\boldsymbol{A}}
\newcommand{\bB}{\boldsymbol{B}}
\newcommand{\bZ}{\boldsymbol{Z}}
\newcommand{\bP}{\boldsymbol{P}}
\newcommand{\bR}{\boldsymbol{R}}
\newcommand{\bS}{\boldsymbol{S}}
\newcommand{\bT}{\boldsymbol{T}}
\newcommand{\bD}{\boldsymbol{D}}
\newcommand{\bL}{\boldsymbol{L}}
\newcommand{\bM}{\boldsymbol{M}}
\newcommand{\bN}{\boldsymbol{N}}
\newcommand{\bG}{\boldsymbol{G}}
\newcommand{\bPi}{\boldsymbol{\Pi}}
\newcommand{\ba}{\boldsymbol{a}}
\newcommand{\bb}{\boldsymbol{b}}
\newcommand{\bu}{\boldsymbol{u}}
\newcommand{\bl}{\boldsymbol{l}}
\newcommand{\bq}{\boldsymbol{q}}
\newcommand{\bk}{\boldsymbol{k}}
\newcommand{\btau}{\boldsymbol{\tau}}
\newcommand{\vareps}{\varepsilon}
\newcommand{\bft}{\textbf{f}}
\newcommand{\bfBO}{\textbf{f}^{\sss{\,(BO)}}}
\newcommand{\bvareps}{\boldsymbol{\varepsilon}}
\newcommand{\bsig}{\boldsymbol{\sigma}}
\newcommand{\bsigBO}{{\bsig}^{\sss{\,(BO)}}}
\newcommand{\bPhi}{\boldsymbol{\Phi}}
\newcommand{\bPsi}{\boldsymbol{\Psi}}
\newcommand{\bvarPhi}{\boldsymbol{\varPhi}}
\newcommand{\bUps}{\boldsymbol{\Upsilon}}
\newcommand{\bvarPsi}{\boldsymbol{\varPsi}}
\newcommand{\bRcal}{\boldsymbol{\Rcal}}
\newcommand{\bDelta}{\boldsymbol{\Delta}}
\newcommand{\bLambda}{\boldsymbol{\Lambda}}
\newcommand{\epsel}{\epsilon^{\scriptscriptstyle{\infty}}}
\newcommand{\bepsel}{\boldsymbol{\epsilon}^{\scriptscriptstyle{\infty}}}
\newcommand{\bbOmega}{\mathbb{\Omega}}

\newcommand{\sss}[1]{\scriptscriptstyle{\text{#1}}}
\newcommand{\hs}{\sss{hs}}

\newcommand{\rscha}{\bRcal}
\newcommand{\rschatrial}{\bRcal}
\newcommand{\phischatrial}{\bvarPhi}
\newcommand{\psischatrial}{\bvarPsi}
\newcommand{\schatrial}{\rscha,\phischatrial}
\newcommand{\Hschatrial}{\Hcal_{\schatrial}}
\newcommand{\HschaR}{\Hcal_{\rschatrial, \phischaR}}
\newcommand{\rhoschatrial}{{\tilde \rho}_{\scriptscriptstyle{\rscha},\scriptscriptstyle{\phischatrial}}}
\newcommand{\rhoschastart}{{\tilde \rho}_{\scriptscriptstyle{\rscha^{(0)}},\scriptscriptstyle{\phischatrial^{(0)}}}}
\newcommand{\rhoR}{\rho_{\rschatrial}}

\newcommand{\scha}{\bRcal,\bvarPhi}
\newcommand{\fBO}{\text{f}^{\sss{\,(BO)}}}
\newcommand{\fscha}{\text{f}^{\Hschatrial}}
\newcommand{\bfscha}{\textbf{f}^{\Hschatrial}}
\newcommand{\bfschaR}{\textbf{f}^{\HschaR}}
\newcommand{\Bigtr}[1]{\textup{Tr}\Bigl[\,#1\,\Bigr]}
\newcommand{\Avg}[2]{\left\langle #1\right\rangle_{#2}}
\newcommand{\Avgschatrial}[1]{\Avg{#1}{ \rhoschatrial}}

\newcommand{\schaR}{\rscha,\bPhi_{\rscha}}
\newcommand{\phischaR}{\bPhi_{\sssrcal}}
\newcommand{\rhoschaR}{\rho_{\schaR}}
\newcommand{\AvgschaR}[1]{\Avg{#1}{ \!\!\rhoschaR}}

\newcommand{\rschaeq}{\bRcal_{\eq}}
\newcommand{\rschahs}{\bRcal_{\hs}}
\newcommand{\schaReq}{\rschaeq,\bPhi(\rschaeq)}
\newcommand{\rhoschaReq}{\rho_{\schaReq}}
\newcommand{\phischaeq}{\bPhi_{\eq}}
\newcommand{\bPhieq}{\bPhi_{\eq}}
\newcommand{\schaeq}{\rschaeq,\phischaeq}
\newcommand{\Hschaeq}{\Hcal_{\schaeq}}
\newcommand{\Hscha}{\Hcal^{\sss{(S)}}}
\newcommand{\Hsscha}{\Hcal^{\sss{(S)}}}
\newcommand{\HS}{\Hcal^{\sss{(S)}}}
\newcommand{\inv}{\scriptscriptstyle{-1}}

\newcommand{\rhoschaeq}{\rho_{\rschaeq,\phischaeq}}

\newcommand{\phischathreeR}{\overset{\sss{(3)}} {\bPhi }{}_{\sssrcal}}
\newcommand{\phischafourR}{\overset{\sss{(4)}} {\bPhi }{}_{\sssrcal}}

\newcommand{\bDR}{\bD_{\sssrcal}}
\newcommand{\DR}{D_{\sssrcal}}

\newcommand{\Phithree}{\overset{\sss{(3)}}{\Phi}}
\newcommand{\bPhithree}{\overset{\sss{(3)}}{\bPhi}}
\newcommand{\Phicent}{\overset{\cent}{\Phi}}
\newcommand{\Dthree}{\overset{\sss{(3)}}{D}}
\newcommand{\bDthree}{\overset{\sss{(3)}}{\bD}}
\newcommand{\DthreeR}{\overset{\sss{(3)}}{D}{}_{\sssrcal}}
\newcommand{\bDthreeR}{\overset{\sss{(3)}}{\bD}{}_{\sssrcal}}
\newcommand{\Dthreeeq}{\overset{\sss{(3)}}{D}{}_{\eq}}
\newcommand{\bDthreeeq}{\overset{\sss{(3)}}{\bD}{}_{\eq}}

\newcommand{\Phifour}{\overset{\sss{(4)}}{\Phi}}
\newcommand{\bPhifour}{\overset{\sss{(4)}}{\bPhi}}
\newcommand{\Dfour}{\overset{\sss{(4)}}{D}}
\newcommand{\bDfour}{\overset{\sss{(4)}}{\bD}}
\newcommand{\DfourR}{\overset{\sss{(4)}}{D}{}_{\sssrcal}}
\newcommand{\bDfourR}{\overset{\sss{(4)}}{\bD}{}_{\sssrcal}}
\newcommand{\Dfoureq}{\overset{\sss{(4)}}{D}{}_{\eq}}
\newcommand{\bDfoureq}{\overset{\sss{(4)}}{\bD}{}_{\eq}}

\newcommand{\Dtwo}{\overset{\sss{(2)}}{D}}
\newcommand{\bDtwo}{\overset{\sss{(2)}}{\bD}}

\newcommand{\bLambdaeq}{\bLambda_{\eq}}
\newcommand{\Lambdaeq}{\Lambda_{\eq}}
\newcommand{\bLambdaR}{\bLambda_{\sssrcal}}
\newcommand{\LambdaR}{\Lambda_{\sssrcal}}

\newcommand{\Dscha}{D^{\sss{(S)}}}
\newcommand{\bDscha}{{\bD}^{\sss{(S)}}}
\newcommand{\Rcaleq}{\Rcal_{\eq}}
\newcommand{\bRcaleq}{\bRcal_{\eq}}
\newcommand{\DF}{D^{\sss{(F)}}}
\newcommand{\bDF}{\bD^{\sss{(F)}}}
\newcommand{\DS}{D^{\sss{(S)}}}
\newcommand{\bDS}{{\bD}^{\sss{(S)}}}
\newcommand{\Eel}{E_{\sss{el}}}
\newcommand{\Fel}{F_{\sss{el}}}
\newcommand{\Sel}{S_{\sss{el}}}
\newcommand{\Sion}{S_{\sss{ion}}}

\newcommand{\Nat}{N_{\sss{a}}}
\newcommand{\Nc}{N_{\sss{c}}}
\newcommand{\Natu}{n_{\sss{a}}}
\newcommand{\kB}{k_{\sss{B}}}

\newcommand{\diag}{\sss{diag}}
\newcommand{\Tc}{T_{\sss{c}}}
\newcommand{\doubledot}{\,\textbf{\text{:}}\,}
\newcommand{\singledot}{\boldsymbol{\cdot}}

\newcommand{\bMinv}{{\bM}^{\boldsymbol{-\frac{1}{2}}}}
\newcommand{\sssrcal}{\scriptscriptstyle{\bRcal}}
\renewcommand{\ss}[1]{\scriptscriptstyle{#1}}
\renewcommand{\Im}{\mathrm{Im}}
\renewcommand{\Re}{\mathrm{Re}}
\newcommand{\static}{\sss{(stat)}}
\newcommand{\bPiB}{\overset{\sss{(B)}}{\bPi}}
\newcommand{\PiB}{\overset{\sss{(B)}}{\Pi}}
\newcommand{\be}{\boldsymbol{e}}
\renewcommand{\bf}{\boldsymbol{f}}
\newcommand{\id}{\sss{id}}
\newcommand{\Id}{\sss{Id}}
\newcommand{\SE}{\sss{SE}}
\newcommand{\OS}{\sss{(os)}}
\newcommand{\se}{\sss{se}}
\newcommand{\pert}{\sss{(pert)}}
\newcommand{\cent}{\sss{(cent)}}
\newcommand{\Tri}{\mathbb{T}}
\renewcommand{\sc}{\sss{sc}}
\renewcommand{\S}{\sss{S}}
\newcommand{\Rlat}{\Rscr_{\sss{lat}}}
\newcommand{\RSlat}{\Rscr^{\sss{(S)}}_{\sss{lat}}}
\newcommand{\RscrS}{\Rscr_{\S}}
\newcommand{\Rscrs}{\Rscr_{\S}}
\newcommand{\SC}{\mathrm{SC}}
\newcommand{\pow}{\sss{pow}}
\newcommand{\ASR}{\sss{(ASR)}}
\newcommand{\PhiASR}{\overset{\ASR}{\Phi}}
\newcommand{\ols}{\overline{s}}
\newcommand{\olbl}{\overline{\bl}}
\newcommand{\Vol}{\Omega_{\sss{Vol}}}

\newcommand{\stress}{P}
\newcommand{\bstress}{\boldsymbol{P}}
\newcommand{\stressbo}{\stress^{\text{(BO)}}}
\newcommand{\bstressbo}{\bstress^{\text{(BO)}}}

\newcommand{\rhoo}{{\rho}}
\newcommand{\hrhoo}{{{\hat \rho}}}

\newcommand{\Avgclassic}[1]{\Avg{#1}{\rho}}
\newcommand{\Avgclassict}[1]{\Avg{#1}{\rho(t)}}
\newcommand{\Avgquantum}[1]{\Avg{#1}{\hat\rho}}
\newcommand{\Avgquantumt}[1]{\Avg{#1}{\hat\rho(t)}}
\newcommand{\Avgclassiceq}[1]{\Avg{#1}{\rhoo}}
\newcommand{\Avgclassiceqz}[1]{\Avg{#1}{\rho^{(0)}}}
\newcommand{\Avgclassicpert}[1]{\Avg{#1}{\rho^{(1)}}}

\newcommand{\Avgquantumeq}[1]{\Avg{#1}{{\hat \rho}^{(0)}}}

\newcommand{\bTheta}{\boldsymbol{\Theta}}
\newcommand{\bC}{\boldsymbol{C}}
\newcommand{\bQ}{\boldsymbol{Q}}
\newcommand{\fharm}{{f^{\bPhi}}}
\newcommand{\bfharm}{\boldsymbol{\fharm}}
\newcommand{\bSigma}{\boldsymbol{\Sigma}}
\newcommand{\Gcal}{{\mathcal G}}
\newcommand{\bGcal}{\boldsymbol{\Gcal}}
\newcommand{\bU}{{\bm U}}
\newcommand{\br}{{\bm r}}

\newcommand{\Hbo}{{\hat H}}
\newcommand{\Htd}{{{\hat H}_{\text{td}}(t)}}

\newcommand{\Vtot}{{V^{(\text{tot})}}}
\newcommand{\ftot}{f^{(\text{tot})}}
\newcommand{\bftot}{\boldsymbol{f}^{(\text{tot})}}
\newcommand{\Vext}{{V^{\text{ext}}}}

\newcommand{\Acal}{{\mathcal A}}
\newcommand{\Bcal}{{\mathcal B}}
\newcommand{\Ramant}{{\Xi}}
\newcommand{\bRamant}{\boldsymbol{\Ramant}}
\newcommand{\bchi}{\boldsymbol{\chi}}

\newcommand{\rhoone}{{{\hat\rho}^{(1)}}}
\newcommand{\Lsc}{ {{L}_\text{sc}}}
\newcommand{\Vsc}{ {\hat{ {V_{\text{sc}}}}^{(1)}}}

\newcommand{\Scal}{{\mathcal{S}}}
\newcommand{\Kcal}{{\mathcal{K}}}
\newcommand{\bv}{\boldsymbol{v}}
\newcommand{\bW}{\boldsymbol{W}}
\newcommand{\Lcal}{{\mathcal L}}
\newcommand{\bKcal}{\boldsymbol{\Kcal}}